\documentclass[10pt]{article}
\pdfoutput=1 

\usepackage{amsmath, amssymb, amsthm} 
\usepackage{natbib}
\usepackage{graphicx}
\usepackage{bm}
\usepackage{booktabs}
\usepackage{url}
\usepackage{hyperref}
\usepackage{color}
\usepackage{lscape}
\usepackage{multirow}
\usepackage{lscape}
\usepackage{fullpage}
\usepackage[a4paper, left=28 mm, right=28 mm, top=24 mm, bottom=39 mm,footskip=12mm, headsep=6mm]{geometry}

\newcommand{\ud}{\mathrm{d}}
\newcommand{\RR}{\mathbb{R}} 
\newcommand{\NN}{\mathbb{N}} 
\newcommand{\PP}{\mathbb{P}}
\newcommand{\EE}{\mathbb{E}}
\newcommand{\bfone}{\mathbf{1}}

\newcommand{\bfX}{\mathbf{X}} 

\newcommand{\bfV}{\mathbf{V}}
\newcommand{\bfv}{\mathbf{v}}
\newcommand{\bfu}{\mathbf{u}} 
\newcommand{\bfU}{\mathbf{U}}
\newcommand{\bfx}{\mathbf{x}} 

\newcommand{\var}{\operatorname{var}}
\newcommand{\VaR}{\operatorname{VaR}} 
\newcommand{\ES}{\operatorname{ES}}

\newcommand{\Psitilde}{\widetilde{\Psi}}
\newcommand{\muMC}{\mu_n}
\newcommand{\muIS}{\widehat{\mu}_n}

\newcommand{\wtilde}{\widetilde{w} }
\renewcommand{\qed}{\ensuremath{\hfill\Box}}

\theoremstyle{plain}
\newtheorem{theorem}{Theorem}

\newtheorem{lemma}[theorem]{Lemma}
\newtheorem{condition}{Condition}

\newtheorem{example}[theorem]{Example}
\newtheorem{algorithm}[theorem]{Algorithm}
\numberwithin{equation}{section}
\numberwithin{theorem}{section}

\begin{document}

\title{An importance sampling approach for copula models in insurance}

\newcounter{savecntr}
\newcounter{restorecntr}

\author{
Philipp Arbenz\thanks{SCOR Global P\&C, General Guisan Quai 26, 8022
Z\"urich, Switzerland \newline  Email: philipp.arbenz@gmail.com} ,
$\quad$Mathieu Cambou\thanks{Institute of Mathematics, Station 8, EPFL, 1015 Lausanne,
Switzerland \newline  Email: mathieucambou@gmail.com}, 
$\quad$Marius Hofert\thanks{Department\ of\ Statistics\ and Actuarial\ Science, 
University of Waterloo, Canada \newline 
Email: marius.hofert@uwaterloo.ca}
}

\date{April 7, 2015}
\maketitle

\begin{abstract}
An importance sampling approach for sampling copula models is introduced. We
propose two algorithms that improve Monte Carlo estimators when the functional of interest depends mainly
on the behaviour of the underlying random vector when at least one of the
components is large. Such problems often arise from dependence models in finance
and insurance. The importance sampling framework we propose is general and can
be easily implemented for all classes of copula models from which sampling is
feasible. We show how the proposal distribution of the two algorithms can be
optimized to reduce the sampling error. In a case study inspired by a typical multivariate 
insurance application, we obtain variance reduction factors between 10 and 30
in comparison to standard Monte Carlo estimators.
  \\
\textit{Key words:} Copula, Dependence models, Importance sampling, Insurance, Risk measure, Tail event\\
\end{abstract}

\section{Introduction}\label{se:introduction}

Many insurance applications, see our motivation Section~\ref{se:Motivation},
lead to the problem of calculating a functional of the form $\EE[\Psi_0(\bfX)]$, where $\bfX=(X_1,\dots,X_d):\Omega\rightarrow\RR^d$ 
is a random vector on a probability space $(\Omega,\mathcal{F},\PP)$ and $\Psi_0:\RR^d\rightarrow\RR$ is a 
measurable function. If the components of $\bfX$ cannot be assumed to be independent, it is popular to model the distribution of $\bfX$ with a copula, such that
\begin{align*}
\PP\left[X_1\leq x_1,\dots,X_d\leq x_d\vphantom{x_{1,}} \right]
= C\left(\vphantom{1_{1_1}}F_{X_1}(x_1),\dots,F_{X_d}(x_d) \right),\qquad \bfx\in\RR^d,
\end{align*}
where $F_{X_j}(x)=\PP[X_j\leq x]$, $j=1,\dots,d$, are the marginal cumulative
distribution functions (cdf) and $C:[0,1]^d\rightarrow[0,1]$ is a copula. A
copula allows one to separate the dependence structure from the marginal
distributions, which is useful for constructing multivariate sto\-ch\-astic
models. We assume the reader to have a basic knowledge on copulas and refer
to~\cite{qrm} or~\cite{nelsen} for introductions.

The usual approach to estimate $\EE[\Psi_0(\bfX)]$ is by Monte Carlo simulation.
In actuarial practice, very often a set of outcomes of $\bfX$ with a low
probability makes a large contribution to $\EE[\Psi_0(\bfX)]$. In this case, importance sampling 
can increase the number of samples lying in this set. Through a weighting approach, 
an unbiased estimator with a reduced variance can be obtained.

Importance sampling for copulas has been investigated by
\cite{Glasserman_Li_2005} and
\cite{HuangSubramanianXuISforCVAR} for the Gauss copula only and
\cite{MarcoBeeAdaptiveCopulaIS} for absolutely continuous copulas.
These papers are inspired by copula models in financial applications and assume 
the copula to be either Gaussian or having a known density. Copulas used in
insurance however often deviate from these assumptions. 

The main contribution
of this paper is the study of importance sampling techniques that do not rely on
a specific copula structure. We consider the case where the functional $\Psi_0$
of interest depends mainly on the behaviour of the random vector $\bfX$ when at least one of
the components is large. Such problems often arise from dependence models in
the realm of finance and insurance, where distorted expectations of heavy tailed
distributions are involved. We propose a new importance sampling framework
for this setup which can be implemented for all classes of copula models from
which sampling is feasible. 

This paper is organized as follows. After motivating our work in
Section~\ref{se:Motivation}, we introduce the importance sampling
approach in Section~\ref{se:ImportanceSampling}.
Section~\ref{se:RejectionAlgorithm} presents a rejection sampling algorithm 
while Section~\ref{se:DirectAlgorithm} presents a direct sampling algorithm. 
For each of them, we expose the sampling of the proposal distribution, the
calculation of the importance sampling weights and we discuss the optimal 
choice of the proposal distribution. Section~\ref{se:rare_event_analysis}
discusses the efficiency of our algorithms in rare event settings.
A case study is given in Section~\ref{se:casestudy} and Section~\ref{se:conclusion} concludes.

\section{Motivation}\label{se:Motivation}

In a copula model, we may write
\begin{align*}
\EE[\Psi_0(\bfX)] = \EE[\Psi(\bfU)]
\end{align*}
where $\bfU=(U_1,\dots,U_d):\Omega\rightarrow\RR^d$ is a random vector with distribution function $C$, $\Psi:[0,1]^d\rightarrow \RR$ is defined as
\begin{align*}
\Psi(u_1,\dots,u_d)= \Psi_0\left(F_{X_1}^{-1}(u_1),\dots,F_{X_d}^{-1}(u_d)\right),
\end{align*}
and $F_{X_j}^{-1}(p) = \inf \{ x\in\RR : F_{X_j}(x)\geq p \}$, for $j=1,\dots,d$.

If $C$ and the margins $F_{X_j}$ are known, we can use Monte Carlo simulation to
estimate $\EE[\Psi(\bfU)]$. For a random sample $\{\bfU_i:i=1,\dots,n\}$ of
$\bfU$, the Monte Carlo estimator of $\EE[\Psi(\bfU)]$ is given by
\begin{align}\label{eq:MCestimator}
\muMC &= \frac{1}{n} \sum_{i=1}^n \Psi(\bfU_i).
\end{align}

In this paper, we consider the case where $\Psi$ is large only when
at least one of its arguments is close to $1$, or equivalently, if at least one of 
the components of $\bfX$ is large. This assumption is inspired by several applications in insurance, as the 
following examples illustrate:
\begin{itemize}
\item The fair premium of a stop loss cover with deductible $T$ is
$\EE\left[\max\left\{\sum_{j=1}^d X_j-T,0\right\}\right]$. The corresponding
functional is $\Psi(\bfu)=\max\left\{\sum_{j=1}^d
F_{X_j}^{-1}(u_j)-T,0\right\}$; see the left hand side of
Figure~\ref{fig:contourplots} for a contour plot of $\Psi$ for two Pareto margins.
\begin{figure}[ht]
{\centering
\scalebox{1}{\includegraphics[width =7.3cm]{Psi_Par_levelplot_lattice.pdf}}
\hspace*{5mm}
\scalebox{1}{\includegraphics[width =7.3cm]{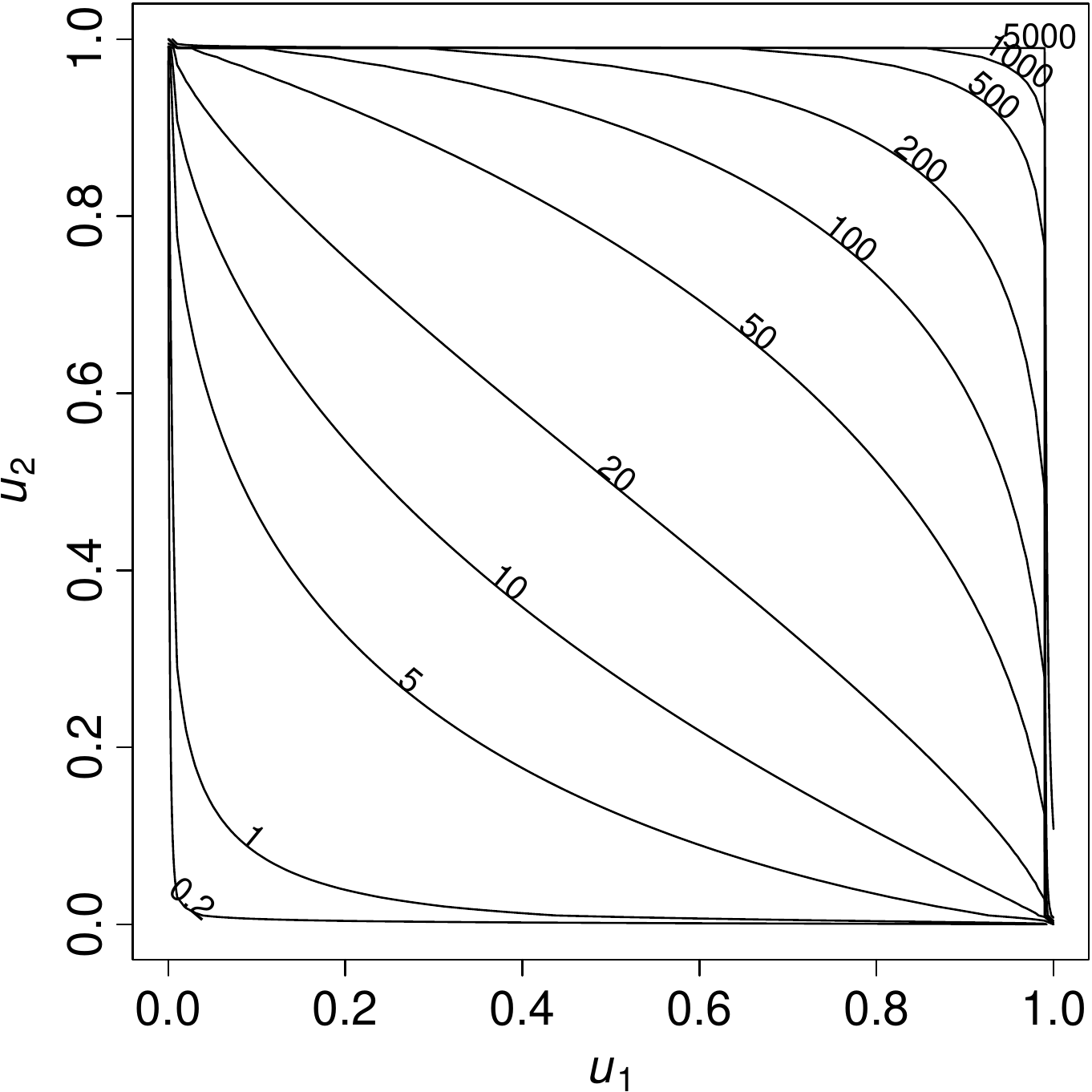}}
\caption{\textit{Left:} Contour lines for the excess function
$\Psi(u_1,u_2)=\max\{F_{X_1}^{-1}(u_1)+F_{X_2}^{-1}(u_2)-10,0\}$, where the margins are Pareto distributed with $F_{X_1}(x)=1-(1+x/4)^{-2}$ and $F_{X_2}(x)=1-(1+x/8)^{-2}$. The grey area indicates where $\Psi$ is zero.
\textit{Right:} Contour lines for the product function
$\Psi(u_1,u_2)=F_{X_1}^{-1}(u_1)F_{X_2}^{-1}(u_2)$, where $X_1\sim \operatorname{LN}(2,1)$ and $X_2\sim \operatorname{LN}(1,1.5)$.}\label{fig:contourplots} }  
\end{figure}
\item Risk measures for an aggregate $S=\sum_{j=1}^d X_j$, such as
Value-at-Risk, $\VaR_\alpha(S)$, or Expected Shortfall, $\ES_\alpha(S)$, $\alpha\in(0,1)$, cannot in general be written as an expectation of type $\EE[\Psi_0(\bfX)]$. However, they are functionals of the aggregate distribution function $F_S(x)=\PP[S\leq x] = \EE[\Psi^{(x)}(\bfU)]$, where $\Psi^{(x)}$ $(x\in\RR)$ is the indicator function 
\begin{align*}
\Psi^{(x)}(\bfu) = \bfone\left\{F_{X_1}^{-1}(u_1)+\dots+F_{X_d}^{-1}(u_d)\leq x\right\}. 
\end{align*}

We can therefore write
\begin{align*}
\VaR_\alpha(S) =\inf\left\{x \in \RR : \EE[\Psi^{(x)}(\bfU)]\geq \alpha\right\},
\qquad
\ES_\alpha(S) = \frac{1}{1-\alpha}\int_\alpha^1 \VaR_u(S)\ud u,
\end{align*}
which depend only on those $x$ for which $\EE[\Psi^{(x)}(\bfU)]\geq\alpha$
holds.
This is determined by the tail behaviour of $S$, which is strongly influenced by
the properties of the copula $C$ when at least one component is close to $1$. Note that capital allocation methods such as the Euler principle for Expected Shortfall behave similarly, see \cite{Tasche2008} and \cite{qrm}, page 260.
\item Computing the covariance (or correlation) of two positive heavy-tailed
random variables $X_1$ and $X_2$ requires the calculation of $\EE[X_1X_2]$. The
implied functional is $\Psi(u_1,u_2)=F_{X_1}^{-1}(u_1)F_{X_2}^{-1}(u_2)$. A
contour plot of $\Psi$ for log-normal (LN) margins is shown in the right hand
side of Figure~\ref{fig:contourplots}. In contrast to the preceding examples,
this $\Psi$ does not only depend on the tail behaviour of $(X_1,X_2)$. However, $\EE[\Psi(\bfU)]$ depends mainly on the copula behaviour when at least one argument is close to 1, as $\Psi$ becomes large in this case.
\end{itemize}

Note that in this framework we follow the convention of~\cite[Remark 2.1]{qrm}
that $\bfX$ refers to a loss and $-\bfX$ to a profit, which is more common
in an actuarial context. One could have equally well worked with the P\&L random
variable $-\bfX$ by changing the area of interest to where components of $\bfX$ are
small.

\section{Importance sampling}\label{se:ImportanceSampling}

The idea behind importance sampling is to sample from a proposal distribution $F_\bfV$ different from the target distribution $C.$ 
The proposal distribution concentrates more samples in the region driving large contributions to $\EE[\Psi(\bfU)]$. 
With a suitable weighting approach, one obtains an unbiased estimator with lower variance.

Suppose the function $\Psi$ under consideration is in the class illustrated above: $\Psi$ is large if at least one of its arguments 
is close to $1$. In this case, a drawback of the estimator $\muMC$ in \eqref{eq:MCestimator} is that, typically, for many of 
the samples $\bfU_i$, none of the components is close to $1$. Therefore, most samples lie in a region of low interest. The estimation 
error of $\muMC$ can thus be large, even if $n$ is large.

Let $\bfV=(V_1,\dots,V_d):\Omega\rightarrow[0,1]^d$ denote a random vector with distribution function $F_\bfV$. We can rewrite the 
integral $\EE[\Psi(\bfU)]$ as
\begin{align}\label{eq:changeofmeasure}
\EE[\Psi(\bfU)]
= \int_{[0,1]^d} \Psi(\bfu)\ud C(\bfu)
= \int_{[0,1]^d} \Psi(\bfu) \frac{\ud C(\bfu)}{\ud F_\bfV(\bfu)}\ud F_\bfV(\bfu)
= \EE\left[\Psi(\bfV)\frac{\ud C(\bfV)}{\ud F_\bfV(\bfV)}\right],
\end{align}
where $\ud C/\ud F_\bfV$ denotes the Radon--Nikodym derivative of $C$ with respect to $F_\bfV$. The Radon--Nikodym derivative exists if and only if the copula $C$ is absolutely continuous with respect to $F_\bfV$. We will provide more details on this issue later in this section. 
If $C$ and $F_\bfV$ are absolutely continuous with densities $c$ and $f_\bfV$
with respect to the Lebesgue measure, the Radon--Nikodym derivative $\ud C/\ud
F_\bfV$ is simply the ratio of the densities $c/f_\bfV$.

For an i.i.d.\,sample $\{\bfV_i:i=1,\dots,n\}$ of $\bfV$, we can define the importance sampling estimator
\begin{align}\label{eq:ISestimator}
\muIS = \frac{1}{n} \sum_{i=1}^n \Psi(\bfV_i) w(\bfV_i),
\end{align}
where $w(\bfV_i)=\ud C(\bfV_i)/\ud F_\bfV(\bfV_i)$ are the sample weights. The
goal is then to find $F_{\bfV}$ such that the variance of $\muIS$
is smaller than the variance of $\muMC$.


In order to define the proposal distribution $F_\bfV,$ we suggest a mixing
approach by taking a weighted average of a multivariate cdf
$C^{[\lambda]}:[0,1]^d\rightarrow [0,1]$ over different values of $\lambda$. Let
$F_\Lambda$ denote the distribution function of a random variable
$\Lambda:\Omega\mapsto[0,1)$. We then define the distribution $F_\bfV$ of $\bfV$
as a mixture of $C^{[\lambda]}$ over the distribution $F_\Lambda$:
\begin{align*}
F_\bfV(\bfu) = 
\int_0^1 C^{[\lambda]}(\bfu)\, \ud F_\Lambda(\lambda), \quad \bfu\in[0,1]^d.
\end{align*}
The distribution $C^{[\lambda]}$ shall be understood as a distorted version of
the copula $C$ that will concentrate samples in specific regions of the sampling
space. These regions will then be parametrized by the value of $\lambda$. More
precisely, we will construct $C^{[\lambda]}$ so that it puts mass only in the
region $[0,1]^d\setminus[0,\lambda]^d$. In the sequel, we will propose two
possible definitions of $C^{[\lambda]}$ that will define two importance sampling algorithms, namely a rejection sampling algorithm in Section~\ref{se:RejectionAlgorithm} and a direct sampling algorithm in
Section~\ref{se:DirectAlgorithm}.

We will see that this mixture approach is natural in order to allow
$C$ to be absolutely continuous with respect to $F_\bfV$. In particular, the absolute
continuity is guaranteed for any copula $C$ if the following condition is satisfied.

\begin{condition}\label{cond:Lambdamass0}
The random variable $\Lambda$ satisfies $\PP[\Lambda=0]>0$.
\end{condition}

In order to obtain a well defined weight function $w(\bfV)$ and an unbiased
estimator $\muIS$, Condition~\ref{cond:Lambdamass0} must be fulfilled. This
condition does not require particular assumptions on $C$. Although it seems
restrictive, we will see that it is also needed to have a consistent estimator
$\muIS$. To that end, we assume Condition~\ref{cond:Lambdamass0} to be satisfied
in what follows.

The construction of the proposal distribution $F_\bfV$ as a
$C^{[\lambda]}$-mixture directly yields a sampling method, as one can draw a
realization of $F_\bfV$ by first drawing $\Lambda \sim F_\Lambda$ and then $\bfV \sim C^{[\Lambda]}$. Therefore, the following algorithm can be used to calculate $\muIS$:
\begin{algorithm}\label{alg:muIScalculation}
Fix $n\in\NN$. For $i=1,\dots,n$, do:
\begin{enumerate}
\item draw $\Lambda_i \sim F_\Lambda$;
\item draw $\bfV_i \sim C^{[\Lambda_i]}$;
\item calculate $w(\bfV_i)$;
\end{enumerate}
Return $\muIS = n^{-1} \sum_{i=1}^n \Psi(\bfV_i) w(\bfV_i)$.
\end{algorithm}

The following lemma establishes consistency and asymptotic normality of the estimator $\muIS$.
\begin{lemma}\label{thm:consistency}
Suppose that $\var[\Psi(\bfU)]<\infty$ and that $w(\,\cdot\,)\leq B$ for some constant $B<\infty$. Then
\begin{enumerate}
\item $\muIS$ converges $\PP$-almost surely to $\mu$;
\item $\sigma^2=\var[\Psi(\bfV)w(\bfV)]<\infty$ and $n^{1/2}(\muIS-\mu)$ converges to $\mathcal{N}(0,\sigma^2)$ in distribution.
\end{enumerate}
\end{lemma}
\textit{Proof.} 
\begin{enumerate}
  \item Since $\EE[\Psi(\bfV)w(\bfV)]=\EE[\Psi(\bfU)]$, consistency follows
  directly from the Strong Law of Large Numbers.
  \item Note that
\begin{align*}
\EE\left[\Psi(\bfV)^2 w(\bfV)^2\right]
= \EE\left[\Psi(\bfU)^2 w(\bfU)\right]
\leq\EE\left[\Psi(\bfU)^2\right]B<\infty,
\end{align*}
where the first equality is justified by a change of measure, see
\eqref{eq:changeofmeasure}. We can immediately deduce asymptotic normality of $\muIS$ by the Central 
Limit Theorem, see, for example, Section 2.4 in~\cite{Durrett}, page 110. \qed
\end{enumerate}

We will later show that under some mild assumptions on $F_\Lambda$, the
weight function will indeed be bounded on $[0,1]$.

\section{A rejection sampling algorithm}\label{se:RejectionAlgorithm}

For this algorithm, we propose $C^{[\lambda]}$ to denote the distribution of $\bfU$ conditioned on the event
that at least one of its components exceeds $\lambda$:
\begin{align}
C^{[\lambda]}(\bfu)
&= \PP\left[U_1\leq u_1,\dots,U_d\leq u_d\,\left\vert\, \max\{U_1,\dots,U_d\}>\lambda \right.\right]
= \PP[U_1\leq u_1,\dots,U_d\leq u_d\,\vert\,\bfU \notin [0,\lambda]^d ] \nonumber \\ \nonumber
&=
\frac{C(\bfu)-C\left(\min\{u_1,\lambda\},\dots,\min\{u_d,\lambda\}\right)}{1-C(\lambda\bfone)},
\end{align}
where $\lambda\bfone = \lambda(1,\dots,1) = (\lambda,\dots,\lambda)\in[0,1)^d$.
Note that $C^{[\lambda]}$ is a copula only if $C(\lambda\bfone)=0$, but
$C^{[\lambda]}$ does not need to be copula for our algorithm to work.  By
putting mass of $\Lambda$ on $(0,1)$, we can put more weight on the region of 
the copula where at least one component is large. For instance, if $F_\Lambda$ is 
discrete and $\PP[\Lambda=0]=\PP[\Lambda=0.9]=0.5$, then 50\% of the samples of $\bfV$ are 
constrained to lie only in $[0,1]^d\setminus [0,0.9]^d$ while the other 50\% of
the samples will lie on $[0,1]^d$. Note that the mass on $[0,1]^d\setminus
[0,0.9]^d$ would then be higher than 50\%. On the other hand, the case
$\PP[\Lambda=0]=1$ yields $F_\bfV=C$.

\subsection{Sampling the proposal
distribution}\label{sse:sampling_rejectionalgorithm}

We shall now describe how samples from $F_\bfV$ can be drawn. As $F_\bfV$ is defined through a mixing distribution, 
drawing a realization from $F_\bfV$ is done by drawing first $\Lambda \sim F_\Lambda$ and then $\bfV \sim C^{[\Lambda]}$, 
see Algorithm~\ref{alg:muIScalculation}. Unfortunately, for well-known copula classes, 
the conditional distribution $C^{[\lambda]}$ is not analytically tractable. We are aware of only one
class of shock copulas, namely Marshall--Olkin copulas, for which it is possible
to sample directly from the conditional distribution $C^{[\lambda]}$. Details  and the
corresponding algorithm are provided in Appendix~\ref{app:AnalyticCtheta}.

However, sampling from $C^{[\lambda]}$ for an arbitrary copula $C$ is
always possible through a rejection algorithm, which is simple to implement but
may be time consuming due to the rejection step. With the following rejection algorithm, 
it is thus possible to draw a sample from $F_\bfV$ for any copula $C$. The only
condition is that it is feasible to draw realizations from both $F_\Lambda$ and $C$. It is not necessary to 
know further properties of $C$, such as its density. The basic idea is to first draw a realization 
$\Lambda$ from $F_\Lambda$ and then iteratively draw realizations from $C$ until one obtains a 
maximum component larger than $\Lambda$.
\begin{algorithm}\label{alg:VsamplingAlgorithm}
To draw one realization of $F_\bfV$:
\begin{enumerate}
\item draw $\Lambda\sim F_\Lambda$;
\item repeatedly draw $\bfV\sim C$ until $\max\{V_1,\dots,V_d\} > \Lambda$;
\item return $\bfV$.
\end{enumerate}
\end{algorithm}
A disadvantage of Algorithm~\ref{alg:VsamplingAlgorithm} is that typically many
samples of $C$ are discarded, because of the acceptance condition in Step 2. 
In practice, there are two important reasons why this approach can be justified
over standard Monte-Carlo. First, the evaluation of $\Psi$ can be numerically
more expensive than sampling from the copula, if, for instance, marginal quantile functions are demanding to compute or if  $\Psi_0$ has no closed form. 
Second, storing a large sample of $\bfU$ in computer memory can be numerically
more expensive than generating it. This case may appear for example in
estimating allocated capital, which requires storing the whole multivariate sample.
In particular in high dimensional problems, memory constraints can be
quite prohibitive.
For illustration, consider the following example: for the calculation of risk capital and risk 
contributions in a setting with heavy tailed marginals, a sample of size
10'000'000 is often not large enough to yield sufficiently small estimation errors. However, in a 1'000-dimensional setting with double-precision 
floating point numbers, this sample would require about 80 gigabytes of memory,
which is more than an average computer currently possesses in terms of RAM.

Algorithm~\ref{alg:VsamplingAlgorithm} may require several realizations from $\bfU$ in order to generate one realization of $\bfV$. The following lemma gives an expression for the expected number of $\bfU$'s for obtaining a realization of $\bfV$.
\begin{lemma}\label{lem:expdraws}
Let $N_\bfV$ denote the number of draws from $C$ necessary to simulate one realization from $F_\bfV$. The expected number of draws is
\begin{align*}
\EE[N_\bfV] = \int_0^1 \frac{1}{1-C(\lambda\bfone)} \ud F_\Lambda(\lambda).
\end{align*}
\end{lemma}
\textit{Proof.} The probability that one draw from $\bfU\sim C$ satisfies $\max\{U_1,\dots,U_d\} > \lambda$ is $\PP[\max\{U_1,\dots,U_d\} > \lambda] = 1-C(\lambda\bfone)$. Therefore, the number of draws necessary to simulate from $C^{[\lambda]}$ for a fixed $\lambda$ is geometrically distributed with parameter $1-C(\lambda\bfone)$ and has expectation $1/[1-C(\lambda\bfone)]$. In order to simulate from $\bfV$, $\Lambda$ is drawn from $F_\Lambda$. Therefore, $\EE[N_\bfV]$ is given by averaging $1/[1-C(\lambda\bfone)]$ over $F_\Lambda$. \qed

Using the Fr\'echet--H\"offding bounds (see Theorem~5.7 in \cite{qrm}), we can give the following bounds for $\EE[N_\bfV]$, which depend only on $F_\Lambda$ and the dimension $d$, independent of the copula $C$.
\begin{theorem}\label{thm:expectationNV} We have 
\begin{align*}
\frac{1}{d}\EE\left[ \frac{1}{1-\Lambda} \right]
\leq \EE[N_\bfV] \leq
\EE\left[ \frac{1}{1-\Lambda} \right].
\end{align*}
\end{theorem}
\textit{Proof.}
Due to the upper Fr\'echet--H\"offding bound, we have $C(\lambda\bfone)\leq \min\{\lambda,\dots,\lambda\}=\lambda$. Hence,
\begin{align*}
\EE[N_\bfV]
&= \int_0^1 \frac{1}{1-C(\lambda\bfone)} \ud F_\Lambda(\lambda)
\leq \int_0^1 \frac{1}{1-\lambda} \ud F_\Lambda(\lambda) =\EE\left[ \frac{1}{1-\Lambda} \right].
\end{align*}
Analogously, due to the lower Fr\'echet--H\"offding bound:
\begin{align*}
\EE[N_\bfV]
\geq \int_0^1 \frac{1}{1-\max\{0,d\lambda -d+1\}} \ud F_\Lambda(\lambda)
= \int_0^1 \max\left\{1,\frac{1}{d(1-\lambda)} \right\} \ud F_\Lambda(\lambda)
\geq \frac{1}{d}\EE\left[ \frac{1}{1-\Lambda} \right]. \qquad \qed
\end{align*}

Due to Theorem~\ref{thm:expectationNV}, the number of draws from $C$ necessary to draw one 
realization from $\bfV$ has a finite expectation if and only if $\EE[ (1-\Lambda)^{-1}] < \infty$. 
Intuitively, this implies that $\Lambda$ should not have mass concentrated near 1 in order to be able 
to use Algorithm~\ref{alg:VsamplingAlgorithm}.

We shall see in the next section that specific choices for the copula $C$ and
for $F_\Lambda$ will allow us to find analytical expressions for $\EE[N_\bfV]$.

\subsection{Calculation of sample weights}\label{sse:weights_rejectionalgorithm}

This section outlines how the weights $w(\bfV_i)$ used in Algorithm~\ref{alg:muIScalculation} 
can be calculated. We first deduce a useful representation.

\begin{theorem}\label{thm:fVrepresentation}
The Radon--Nikodym derivative $w(\bfu)=\ud C(\bfu)/\ud F_\bfV(\bfu)$ can be written as
\begin{align*}
w(\bfu)
= \left( \int_0^{\max\{u_1,\dots,u_d\}} \frac{1}{1-C(\lambda\bfone)} \ud F_\Lambda(\lambda) \right)^{-1}.
\end{align*}
\end{theorem}
\textit{Proof.} Due to the Leibnitz integral rule, we have $\ud F_\bfV(\bfu) = \int_0^1  \ud C^{[\lambda]}(\bfu) \ud F_\Lambda(\lambda)$. From the definition of $C^{[\lambda]}$, we can deduce the differential
\begin{align*}
\ud C^{[\lambda]}(\bfu) =
\begin{cases}
0, & \bfu \in [0,\lambda]^d, \\
\frac{\ud C(\bfu)}{1-C(\lambda\bfone)}, & \text{ otherwise. }
\end{cases}
\end{align*}
Using both identities, we obtain
\begin{align*}
\ud F_\bfV(\bfu) = \ud C(\bfu)
\int_0^1 \frac{\bfone\left\{\lambda \leq \max\{u_1,\dots,u_d\}\right\}}{1-C(\lambda\bfone)}\ud F_\Lambda(\lambda),
\end{align*}
leading to the desired result. \qed

The efficiency of our approach comes from the fact that the term $\ud C(\bfu)$ does not appear in $w(\bfu)$. For instance, if $C$ is absolutely continuous with respect to the Lebesgue measure, the density of $C$ does not have to be evaluated to calculate $w(\bfu)$. 
This is in an advantage in comparison to most other importance sampling
algorithms, for which the existence of the density of $C$ is required.

In order to simplify the notation, let $\wtilde(t):[0,1]\rightarrow[0,\infty)$ be defined as
\begin{align*}
\wtilde(t)= \left( \int_0^t \frac{1}{1-C(\lambda\bfone)} \ud F_\Lambda(\lambda)\right)^{-1},
\end{align*}
such that $w(\bfu)=\wtilde(\max\{u_1,\dots,u_d\})$.

\begin{lemma}
Under Condition~\ref{cond:Lambdamass0}, $\wtilde$ is bounded from above by
$\PP[\Lambda=0]^{-1}$ on $[0,1].$
\end{lemma}
\textit{Proof.} 
Since $C(\lambda\bfone),\lambda\in[0,1]$, the diagonal section of the copula $C$
and the distribution function $F_\Lambda$ are both increasing functions, the
weight function $\wtilde(t)$ is decreasing on $[0,1]$, it is therefore bounded above by $\wtilde(0)=\PP[\Lambda=0]^{-1}<\infty.$ \qed

As a consequence, Condition~\ref{cond:Lambdamass0} is not only sufficient to obtain existence of the weights, but it also guarantees that they are bounded. In virtue of Lemma~\ref{thm:consistency}, 
this is needed for consistency and asymptotic normality of the importance
sampling estimator.

For general $C$ and $F_\Lambda$, the evaluation of the weight function $\wtilde$ can be demanding. 
In general, numerical integration schemes could be used. To circumvent these problems, we present two cases in 
which the evaluation of $\wtilde$ is straightforward.
Section~\ref{sse:discreteFtheta_weights_rejectionalgo} illustrates the case in which $F_\Lambda$ is discrete. 
In Section~\ref{sse:continuousFtheta_weights_rejectionalgorithm}, we assume that the  copula $C$ lies in a
large class of copulas satisfying a polynomial condition on the diagonal. For this class, there is a specific choice 
of $F_\Lambda$ which leads to an analytical expression for $\wtilde$.

\subsubsection{Discrete
$F_\Lambda$}\label{sse:discreteFtheta_weights_rejectionalgo}

This section shows that in the case of a discrete $F_\Lambda$, calculating
$\wtilde(t)$ is fast and can easily be implemented. To this end, suppose
$F_\Lambda$ is discrete with a finite number $n_\Lambda$ of atoms:
\begin{align*}
\PP[\Lambda = x_k]=p_k, k=1,\dots, n_\Lambda,
\sum_{k=1}^{n_\Lambda}p_k=1, p_1>0, \text{ and } 0= x_1 <\dots< x_{n_\Lambda} <
1.
\end{align*}
Note that Condition~\ref{cond:Lambdamass0} is satisfied. In this case,
$\wtilde$ can be written as a step function
\begin{align}\label{eq:discreteweightfunction}
\wtilde(t)
=\left( \sum_{k=1}^{n_\Lambda} \frac{\bfone\{x_k\leq t\}}{1-C(x_k\bfone)}p_k \right)^{-1}.
\end{align}
In order to evaluate $\wtilde(t)$, it is sufficient to calculate (or approximate) $C(x_k\bfone)$ for $k=1,\dots,n_\Lambda$. 
These values must be calculated only once for the whole sample. This approach with a discrete $F_\Lambda$ can be used for 
any copula $C$. For $\EE[N_\bfV]$, we obtain the explicit
expression
\begin{align*}
\EE[N_\bfV] = \sum_{k=1}^{n_\Lambda}\frac{p_k}{[1-C(x_k\bfone)]}.
\end{align*}

\subsubsection{Continuous
$F_\Lambda$}\label{sse:continuousFtheta_weights_rejectionalgorithm}

For continuous $F_\Lambda$, the weight function $\wtilde$ can in general only be
calculated numerically. In the following, we assume that both $C$ and
$F_\Lambda$ are of a special polynomial form, which leads to an explicit
$\wtilde$.
Suppose that $C$ behaves as a monomial on its diagonal:
\begin{align*}
C(u \mathbf{1})=u^\alpha,
\quad 0\leq u \leq 1.
\end{align*}
Due to the Fr\'echet--H\"offding bounds, $\alpha$ must satisfy $1\leq\alpha\leq
d$. This class of copulas is quite large. The following list shows some popular copula families satisfying this condition.
\begin{itemize}
\item Marshall--Olkin copulas as proposed in Example~\ref{ex:MOC} of
Appendix~\ref{app:AnalyticCtheta}. The corresponding exponent is $\alpha =
\sum_{j=1}^m \min_{i:j\in I_i} (s_j/\widetilde{s}_i)$.
\item Sibuya copulas, as defined in \cite{Sibuya2011}, for which the default rate process is a homogeneous Poisson process.
\item Extreme value copulas with a Pickands dependence function $A$. The
corresponding exponent is $\alpha=d A(1/d,\dots,1/d)$; see Section~7 in
\cite{qrm} for a definition of extreme value copulas. Note that this class
contains the well-known Gumbel copula, for example.
\end{itemize}

Apart from the copula $C$, we also make some specific assumptions about
$F_\Lambda:[0,1]\rightarrow[0,1]$. Suppose that
\begin{align*}
F_\Lambda(\lambda)=(1-\gamma)+\gamma\left(1-(1-\lambda^{\alpha})^{\beta}\right),
\quad \beta>1,\,0\leq\gamma\leq1.
\end{align*}
The parameter $\alpha$ is given by the exponent of the copula diagonal, so
cannot be chosen freely. Furthermore,
$F_\Lambda$ has an atom of weight $1-\gamma$ at zero. This distribution is similar to the distribution of \cite{KumaraswamyDistribution}. In this case, the weight function can easily be calculated as
\begin{align}\label{eq:continuousweights_rejection}
\widetilde{w}(t) = \left(1-\gamma+\gamma\beta\int_0^t \alpha\lambda^{\alpha-1}(1-\lambda^\alpha)^{\beta-2}\ud \lambda\right)^{-1}
= \frac{\beta-1}{\beta-1+\gamma\left(1-\beta(1-t^\alpha)^{\beta-1}\right)}.
\end{align}
As $\EE[N_\bfV]=1/\wtilde(1)$ (c.f.\ Lemma~\ref{lem:expdraws}), we obtain an
explicit expression for $\EE[N_\bfV]$:
\begin{align}
\EE[N_\bfV]=1+\frac{\gamma}{\beta-1}.
\end{align}
In order for Condition~\ref{cond:Lambdamass0} to be satisfied, we assume $\gamma<1$. In fact, 
using properties of the hypergeometric function, it is possible to show that for $\gamma=1$, 
the weight function is unbounded and the variance of the weights $\var[w(\bfV)]$ is always infinite.

There are many copula classes which have an explicit diagonal. For instance, the Clayton family has a diagonal
$C(t\bfone)=(dt^{-\theta}-d+1)^{-1/\theta}$ for some $0<\theta<\infty$. For
future research, we may point out that it would be interesting to find
``conjugate'' $F_\Lambda$ for copulas that also allow for an explicit form of
$\wtilde(\cdot)$.

\subsection{Optimal proposal
distribution}\label{sse:optimalthetadistribution_rejectionalgorithm}

This section gives an approach to calibrate the distribution $F_\Lambda$ to
the problem at hand. The basic idea is to choose the proposal distribution
$F_\bfV$ in such a way that $\muIS$ has a smaller variance than $\muMC$. In our
case, this reduces to optimally choosing the distribution $F_\Lambda$. In
general, $F_\Lambda$ must have an atom at $0$ in order to satisfy Condition~\ref{cond:Lambdamass0}.
If Algorithm~\ref{alg:VsamplingAlgorithm} is used for sampling, we also need to fulfill the constraint 
that $\EE[1/(1-\Lambda)]$ is not too large, and, in particular, finite.

Zero variance (i.e., no estimation error) would be obtained for $\muIS$ if
\begin{align}\label{eq:optimalcondition}
\Psi(\bfu)w(\bfu) = \EE[\Psi(\bfU)], \quad \bfu\in[0,1)^d,
\end{align}
see Section 4.1 in \cite{AsmussenGlynn2007}, page 128. 
This choice is obviously not possible as $\EE[\Psi(\bfU)]$ is unknown. To obtain
a small variance, we should choose $\Lambda$ such that $w(\bfu)^{-1}$ is approximately proportional to $\Psi(\bfu)$. Due to Theorem~\ref{thm:fVrepresentation}, we may write this relation as
\begin{align}\label{eq:optimalH1}
 K \int_0^{\max\{u_1,\dots,u_d\}} \frac{1}{1-C(\lambda\bfone)} \ud F_\Lambda(\lambda) \approx\Psi(\bfu),
\end{align}
for some unknown constant $K\in\RR_{+}$. In order to obtain a tractable
optimization scheme, we use our assumption that $\Psi(\bfu)$ is
large if at least one of its components is large, namely
\begin{align}\label{eq:PsiLargeApprox}
\Psi(\bfu) \approx \Psi\bigl(\max\{u_1,\dots,u_d\}\bfone\bigr).
\end{align}
Plugging \eqref{eq:PsiLargeApprox} into \eqref{eq:optimalH1}, we obtain
\begin{align}\label{eq:optimalreducedeq_rejection}
K \int_0^{t} \frac{1}{1-C(\lambda\bfone)} \ud F_\Lambda(\lambda)
\approx \Psi(t\mathbf{1}), \quad t\in[0,1].
\end{align}
In the following, we propose methods to calibrate $F_\Lambda$ such that the approximate relation 
\eqref{eq:optimalreducedeq_rejection} is satisfied. 
We illustrate this calibration with the two choices for $F_\Lambda$ being
discrete and continuous.

\subsubsection{Discrete
$F_\Lambda$}\label{sse:Optim_discrete_optimalthetadistribution_rejectionalgorithm}

In the discrete case, as defined in Section~\ref{sse:discreteFtheta_weights_rejectionalgo}, specifying the distribution $F_\Lambda$ 
reduces to setting the atoms $x_k$ and their weights $p_k=\PP[\Lambda=x_k]$ for $k=1,\dots,n_\Lambda$. By plugging $F_\Lambda$ 
into~\eqref{eq:optimalreducedeq_rejection}, we obtain
\begin{align}\label{eq:linearconditiononTheta_rejection}
K\sum_{k=1}^{n_\Lambda} \frac{\bfone\{x_k\leq t\}p_k}{1-C(x_k\bfone)}
 \approx\Psi(t\bfone), \quad t\in[0,1).
\end{align}
We propose to set the $p_k$'s by enforcing equality to hold in
\eqref{eq:linearconditiononTheta_rejection} only for
$t=x_1,\dots,x_{n_\Lambda}$. By assuming without loss of generality, that 
$x_k<x_{k+1}$ for all $k$, Equation \eqref{eq:linearconditiononTheta_rejection} leads to
\begin{align*}
 K\sum_{l=1}^k \frac{1}{1-C(x_l\bfone)}p_l =\Psi(x_k\bfone), \quad k=1,\dots,n_\Lambda.
\end{align*}
This yields a triangular linear system of equations which can be easily solved 
with the following algorithm; we propose to choose the $x_k$'s on a finite
logarithmic grid becoming denser towards $1$.
\begin{algorithm}\label{alg:ThetaDistribution_rejection} $\,$
\begin{enumerate}
\item Choose $n_\Lambda\in\NN$;
\item define $x_k=1-(1/2)^{k-1}$, $k=1,\dots,n_\Lambda$;
\item define $\widetilde{p}_1=\Psi(0,\dots,0)$ and $\widetilde{p}_k = \left(\Psi(x_k\bfone)-\Psi(x_{k-1}\bfone)\right)(1-C(x_k\bfone))$, for $k=2,\dots,n_\Lambda$;
\item define $p_k=\widetilde{p}_k/(\sum_l\widetilde{p}_l)$.
\end{enumerate}
\end{algorithm}

The use of powers of $1/2$ to set the $x_k$'s is arbitrary; any other factor in
$(0,1)$ can be used instead. In numerical experiments, the impact of this choice was in general small, 
as the calculated $p_k$ change accordingly.

In the following situations, Algorithm~\ref{alg:ThetaDistribution_rejection}
may fail:
\begin{itemize}
\item if $p_1=0$, then $F_\Lambda$ does not satisfy Condition~\ref{cond:Lambdamass0};
\item if $t\mapsto\Psi(t\bfone)$ is not monotone, then Algorithm~\ref{alg:ThetaDistribution_rejection} results in some of the $p_k$'s being negative;
\item if the function $\Psi$ does not attain a finite value at $(0,\dots,0)$.
\end{itemize}

Since $n_\Lambda<\infty$, the condition $\EE[1/(1-\Lambda)]<\infty$ is
automatically satisfied. Of course, one could also use discrete
distributions for $\Lambda$ supported by infinitely many points. However, in
experiments analogous to the case study presented in Section~\ref{se:casestudy}, this has led to waiting times
$\EE[N_\bfV]$ becoming large without providing additional accuracy when using
rejection sampling.

\subsubsection{Continuous
$F_\Lambda$}\label{se:Optim_continuous_optimalthetadistribution_rejectionalgorithm}

In the continuous case, as defined in Section~\ref{sse:continuousFtheta_weights_rejectionalgorithm}, the
optimization unfortunately cannot be done as easily and explicitly as for the
discrete case. By putting $F_\Lambda$, see Equation
\eqref{eq:continuousweights_rejection}, into
\eqref{eq:optimalreducedeq_rejection}, we obtain
\begin{align}\label{eq:continuous_lambda_tosolve}
K\left(1+\frac{\gamma\left(1-\beta(1-t^\alpha)^{\beta-1}\right)}{\beta-1}\right)
\approx \Psi(t\mathbf{1}), \quad t\in[0,1].
\end{align}
In order to optimize $F_\Lambda$, we would need to find parameters $K\in\RR$,
$\gamma\in(0,1)$ and $\beta>1$ which minimize some distance between the left and
right hand side of \eqref{eq:continuous_lambda_tosolve}. As distance function,
one can for instance use the quadratic norm. This minimization can be
implemented through standard numerical minimization procedures. Recall that
$\alpha$ is fixed through the copula's diagonal. In order to have $\EE[N_\bfV]$
not excessively high, one might want to impose a further parameter constraint by bounding $\EE[N_\bfV]=1+\gamma/(\beta-1)$.

\section{A direct sampling algorithm}\label{se:DirectAlgorithm}
As noted in the previous section, the rejection sampling algorithm may lead to
large sampling time due to the rejection step. This step was necessary due to
the complexity of the conditioning event in the definition of $C^{[\lambda]}$.
We now consider
\begin{align}\label{eq:copula_direct_algorithm}
C^{[\lambda]}(\bfu)&=d^{-1} \sum_{i=1}^d \PP\left[U_1\leq u_1,\dots, U_d\leq
u_d\,|\,U_i>\lambda\right]\\
&=d^{-1}\sum_{i=1}^d
\frac{C(\bfu)-C\left(u_1,\dots,u_{i-1},\min\{u_i,\lambda\},u_{i+1},\dots,u_d\right)}{1-\lambda}\nonumber,\quad
\bfu\in[0,1]^d.
\end{align}
This distribution only involves the conditional copula where the conditioning
event is only on one element of the random vector $\bfU$. This will have the
practical advantage that a direct sampling algorithm, i.e. with no rejection
step, can be provided.

\subsection{Sampling the proposal
distribution}\label{sse:sampling_directalgorithm}

Let us denote $C_{u_k}$ the conditional copula given that the $k$-th component
equals $u_k$, that is
\begin{align*}
C_{u_k}(u_1,\dots,u_{k-1},u_{k+1},\dots,u_d)=\PP[U_1\leq u_1,\dots,U_{k-1}\leq
u_{k-1},U_{k+1}\leq u_{k+1},\dots,U_{d}\leq u_{d}\,|\,U_k = u_k].
\end{align*}
Sampling from $F_\bfV$ can then be performed using the following algorithm.

\begin{algorithm}\label{alg:VconditionalSamplingAlgorithm}
To draw one realization of $F_\bfV$:
\begin{enumerate}
\item draw $\Lambda\sim F_\Lambda$;
\item draw $I\in\{1,\dots,d\},$ with $\PP[I=i]=d^{-1}$;
\item draw $V_I\sim U(\Lambda,1)$;
\item draw $(V_1,\dots,V_{I-1},V_{I+1},\dots,V_d)\sim C_{V_I}$;
\item return $\bfV=(V_1,\dots,V_d)$.
\end{enumerate}
\end{algorithm}

The main advantage of this algorithm is that it does not reject any sample and
as a consequence, in contrast to Algorithm
\ref{alg:VsamplingAlgorithm}, its run time does not depend on the distribution
$F_\Lambda.$ In addition, one can show that using a rejection algorithm 
for producing samples from~\eqref{eq:copula_direct_algorithm} would yield an
expected waiting time of $\EE[(1-\Lambda)^{-1}]$, which would be higher than the
expected waiting time of the rejection sampling presented in
Section~\ref{se:RejectionAlgorithm}, see Theorem~\ref{thm:expectationNV}. This
justifies the fact that we propose two specific distributions of
$C^{[\lambda]}$ for each of these two algorithms.

In Step 4 of Algorithm~\ref{alg:VconditionalSamplingAlgorithm}, a sampling
algorithm for the conditional copula $C_{u_k}$, where $k$ can be any of the $d$ components, 
is required. Depending on the form of the copula $C_{u_k}$, efficient sampling
algorithms may be available, see for instance Examples~\ref{ex:FGM} and~\ref{ex:Frank} below, 
or one can use the conditional distribution method. Note that the conditional
distribution method is applied, for example, for sampling vine
copulas; see the \texttt{VineCopula} \textsf{R} package.

Along the lines of~\cite{embrechtslindskogmcneil2003}, we then propose the following general
algorithm to sample from $C_{u_k}$.

\begin{algorithm}\label{alg:conditionalcopulaSampling}
Given $u_k\in\mathbb{R}$, to draw one realization of $C_{u_k}$, do:
\begin{enumerate}
  \item draw $\bfU'=\left(U_1',\dots, U_{k-1}',U_{k+1}',\dots,U_d'\right)\sim
  U(0,1)^{d-1}$;
  \item set 
  \begin{align*}
  U_1 &= C^{-1}(U_1'\,|\,u_k)\\
  \vdots & \\ 
  U_{k-1} &= C^{-1}(U_{k-1}'\,|\,U_1,\dots, U_{k-2},u_k)\\
  U_{k+1} &= C^{-1}(U_{k+1}'\,|\,U_1,\dots, U_{k-2},U_{k-1},u_k)\\
  \vdots & \\
  U_d & = C^{-1}(U_d'\,|\,U_1,\dots,U_{k-1},u_k,U_{k+1},\dots, U_{d-1})
  \end{align*}
  \item return $(U_1,\dots, U_{k-1},U_{k+1},\dots, U_d)$.
\end{enumerate}
\end{algorithm}

Following Theorem 2.27 and Remark 2.29 in~\cite{schmitzthesis}, we have that for
$k>j$
\begin{align}\label{eq:conditional_copulas_differentials}
C(u_j|u_1,\dots, u_{j-1}, u_k) = \frac{D_{1,\dots,j-1,k}
C_{1,\dots,j-1,j,k}(u_1,\dots,u_{j-1},u_j,u_k)}{D_{1,\dots,j-1,k}
C_{1,\dots,j-1,k}(u_1,\dots,u_{j-1},u_k)},
\end{align} 
which simplifies to 
\begin{align*}
C(u_j|u_1,\dots, u_{j-1}) = \frac{D_{1,\dots,j-1}
C_{1,\dots,j-1,j}(u_1,\dots,u_{j-1},u_j)}{D_{1,\dots,j-1}
C_{1,\dots,j-1}(u_1,\dots,u_{j-1})},
\end{align*}
whenever $k<j$. Here, $D_{1,\dots,j,k}$ denotes the partial derivatives with
respect to the components ${1,\dots,j,k}$ and $C_{1,\dots,j,k}$ denotes the
copula corresponding to the distribution of these components. In general, tractable inverses of the
conditional distributions~\eqref{eq:conditional_copulas_differentials} are
not always available, and numerical root-finding would need to be applied.
However, there are cases where one can derive explicitly such inverses, see,
e.g., Example~\ref{ex:conddistribClayton}. In consequence, although this
algorithm does not involve a rejection step, it may require more implementation effort. 

\begin{example}[Direct sampling of Farlie--Gumbel--Morgenstern
copula]\label{ex:FGM} The Farlie--Gumbel--Morgenstern (FGM) copula is defined by 
\begin{align*}
C^{\theta}(\bfu) = \prod_{i=1}^d u_i \left(1+\theta \prod_{j=1}^d
(1-u_j)\right),\quad \bfu\in\mathbb{R}^d,
\end{align*}
with $\theta\in[-1,1]$, see, e.g., \cite{FGMGenest2011}. This copula is a
special form of the more general Eyraud--Farlie--Gumbel--Morgenstern copula, see
page 19 in~\cite{jaworskidurantehaerdlerychlik2010}. It is easily seen that
\begin{align*}
\frac{\partial}{\partial u_k} C^{\theta}(\bfu)&=\prod_{i=1, i\neq
k}^{d}u_i\left(1+\theta (1-2 u_k)\prod_{i=1, i\neq k}^{d}(1-u_i)\right)\\
&= C^{\theta(1-2u_k)}(u_1,\dots, u_{k-1},u_{k+1},\dots, u_d),
\end{align*}
where $C^{\theta(1-2u_k)}$ is a FGM copula with parameter
$\theta(1-2u_k)\in[-1,1]$. As a consequence, sampling from $C_{u_k}^\theta$ is
reduced to sampling from $C^{\theta(1-2u_k)}$. To this end, the conditional
distribution method can be used. Producing a sample
$\bfU\sim C^\theta$ can indeed be reduced to drawing $\bfU'\sim U(0,1)^d$ and
setting $U_1=U_1',\dots,U_{d-1}=U_{d-1}'$, and 
\begin{align*}
U_d = \frac{2U_d'}{1+W+\sqrt{(1+W)^2 - 4 W U_d'}},
\end{align*}
where $W=\theta\prod_{j=1}^{d-1}(1-2U_j')$, see Section 8.7.12
in~\cite{FinancialEngineeringRemillard} for more details.
\end{example}

\begin{example}[Direct sampling of Frank
copula]\label{ex:Frank}
According to Section 6 in~\cite{MesfiouiQuessy2007}, if
$C(\bfu)=\psi\left(\psi^{-1}(u_1)+\dots+\psi^{-1}(u_d)\right)$ is a
$d$-dimensional Archimedean copula with generator $\psi$, then the $(d-1)$-dimensional copula
$\widetilde{C}$ of the multivariate distribution $C_{u_k}$ is
also Archimedean, with generator 
\begin{align*}
\psi_{u_k}(t)=\frac{\psi'(t+\psi^{-1}(u_k))}{\psi'(\psi^{-1}(u_k))},\quad
t\in[0,\infty].
\end{align*} 
This can be used to show that if $C$ is a Frank copula with parameter
$\alpha\in\mathbb{R}$ and generator
$\psi_\alpha(t)=-\alpha^{-1}\log(1-(1-e^{-\alpha})e^{-t})$, then $C_{u_k}$ can
be modeled by a multivariate distribution with copula of type Ali--Mikhail--Haq
with parameter $\theta(\alpha,u_k)=1-e^{-\alpha u_k}$, generator
\begin{align}\label{eq:ex_amh_generator}
\psi_{\theta(\alpha,u_k)}(t)=\frac{1-\theta(\alpha,u_k)}{e^t -
\theta(\alpha,u_k)}
\end{align}
and marginal distributions that have quantile functions
\begin{align}\label{eq:ex_margins}
F_{\alpha,
u_k}^{-1}(u)=-\frac{1}{\alpha}\log\left(\frac{e^{-\alpha}-1}{1+e^{-\alpha
u_k}(u^{-1}-1)}+1\right), \quad u\in[0,1].
\end{align}
In consequence, sampling from $C_{u_k}$ is reduced to sampling from a
Ali--Mikhail--Haq copula with generator~\eqref{eq:ex_amh_generator}, for example
using the fast Marshall--Olkin algorithm, see Sections 2.4 and 2.5
in~\cite{mariusphdthesis}, and then applying the quantile
function~\eqref{eq:ex_margins} to the copula sample. In a similar fashion,
if $C$ is Archimedean such that $C_{u_k}$ is easy to sample with the
Marshall--Olkin algorithm (many examples and techniques are known), and,
additionally, the marginal distributions are easy to invert, then one
obtains a fast sampling technique for Step 4 in Algorithm~\ref{alg:VconditionalSamplingAlgorithm}.
\end{example}

\begin{example}[Conditional distribution method for Clayton copula]\label{ex:conddistribClayton}
The Clayton copula is defined by 
\begin{align*}
C^{\theta}(\bfu) = \left(1+\sum_{i=1}^d
(u_i^{-\theta}-1)\right)^{-1/\theta},\quad \bfu\in\mathbb{R}^d,
\end{align*}
with $\theta>0$. Using~\eqref{eq:conditional_copulas_differentials}, one can
show that
\begin{align*}
C^{\theta(-1)}(u_j'|u_1,\dots, u_{j-1}, u_k) =
\left(1+\left(1-(j-1)+\sum_{k=1}^{j-1}u_k^{-\theta}\right)
\left((u_j')^{-\frac{1}{j-1+1/\theta}}-1\right)\right)^{-1/\theta},
\end{align*}
which allows one to easily implement
Algorithm~\ref{alg:conditionalcopulaSampling}.
\end{example}

\subsection{Calculation of sample weights}\label{sse:weights_directalgorithm}

As for the rejection sampling approach, we derive a representation for the
weights $w(\bfV_i)$ used in Algorithm~\ref{alg:VconditionalSamplingAlgorithm}.

\begin{theorem}\label{thm:fVrepresentation_directalgorithm}
The Radon--Nikodym derivative $w(\bfu)=\ud C(\bfu)/\ud F_\bfV(\bfu)$ can be written as
\begin{align*}
w(\bfu)
= \left( d^{-1}\sum_{i=1}^d\int_0^{u_i}
\frac{1}{1-\lambda} \ud F_\Lambda(\lambda) \right)^{-1}.
\end{align*}
\end{theorem}
\textit{Proof.} Noting that 
\begin{align*}
\ud C^{[\lambda]}(\bfu) = \frac{\ud C(\bfu)}{d(1-\lambda)}\sum_{i=1}^d
\bfone\{u_i>\lambda \},
\end{align*}
we proceed similarly as in the proof of
Theorem~\ref{thm:fVrepresentation}.\qed

As in the rejection sampling algorithm, we note that $\ud C(\bfu)$ does not
appear in $w(\bfu)$, so that the existence of the density of $C$ is not a
requirement for the derivation of the weights. In order to insure consistency
and asymptotic normality of the importance sampling estimator, we shall
also check the boundedness of the weight function.

\begin{lemma}
Under Condition~\ref{cond:Lambdamass0}, the weight function $w$ is bounded from
above by $\PP[\Lambda=0]^{-1}$ on $[0,1].$
\end{lemma}
\textit{Proof.} 
We note that $w(\bfu)$ is decreasing in all components. Hence, it is bounded above by 
$w(0,\dots,0)=\PP[\Lambda=0]^{-1}<\infty.$ \qed

For general $F_\Lambda$, the evaluation of the weight function $w$ could
be demanding. In general, numerical integration schemes could be used. To
circumvent these problems we propose to use the same setup for $F_\Lambda$ as in
Section~\ref{se:RejectionAlgorithm}, i.e., a discrete case 
and a continuous case.

\subsubsection{Discrete $F_\Lambda$}\label{sse:discreteFtheta_weights_directalgorithm}

If $F_\Lambda$ is discrete such that $\PP[\Lambda = x_k]=p_k$, $p_1>0$,
$k=1,\dots, n_\Lambda,$ $0= x_1 <\dots< x_{n_\Lambda} < 1$, then $w$ can
be written as
\begin{align}\label{eq:discreteweightfunction_directalgorithm}
w(\bfu)
=d\left(\sum_{i=1}^{d}\sum_{k=1}^{n_\Lambda} \frac{\bfone\{x_k\leq
u_i\}}{1-x_k}p_k \right)^{-1}.
\end{align}

\subsubsection{Continuous
$F_\Lambda$}\label{sse:continuousFtheta_weights_directalgorithm}

Taking the cdf of $\Lambda$ as
\begin{align*}
F_\Lambda(\lambda)=(1-\gamma)+\gamma\left(1-(1-\lambda)^{\beta}\right),
\quad \beta>1,\,0\leq\gamma<1,
\end{align*}
gives, for any copula $C$, the following closed form for the weights 
\begin{align}\label{eq:continuousweights_direct}
w(\bfu) =
\frac{\beta-1}{\beta-1+\gamma-\gamma\beta d^{-1}\sum_{i=1}^d
(1-u_i)^{\beta-1}}.
\end{align}
Note that we do not need any restriction on the copula diagonal, in contrast
to Section~\ref{sse:continuousFtheta_weights_rejectionalgorithm}

\subsection{Optimal proposal
distribution}\label{se:optimalthetadistribution_directalgorithm}

To obtain a small variance, we should choose $\Lambda$ such that $w(\bfu)^{-1}$ is approximately 
proportional to $\Psi(\bfu)$. Due to Theorem~\ref{thm:fVrepresentation_directalgorithm}, we may write this relation as
\begin{align}\label{eq:optimalH1_directalgorithm}
 K d^{-1} \sum_{i=1}^d \int_0^{u_i}
 \frac{1}{1-\lambda} \ud F_\Lambda(\lambda) \approx\Psi(\bfu),
\end{align}
for some unknown constant $K\in\RR_{+}$. As per the rejection sampling
approach, we shall restrict the calibration to the diagonal in order to obtain
a tractable optimization scheme and we use our assumption that $\Psi(\bfu)
\approx \Psi\bigl(\max\{u_1,\dots,u_d\}\bfone\bigr)$.
Therefore~\eqref{eq:optimalH1_directalgorithm} reduces to
\begin{align}\label{eq:optimalreducedeq}
K \int_0^{t} \frac{1}{1-\lambda} \ud F_\Lambda(\lambda)
\approx \Psi(t\mathbf{1}), \quad t\in[0,1].
\end{align}
In the following, we propose methods to calibrate $F_\Lambda$ such that the approximate 
relation \eqref{eq:optimalreducedeq} is satisfied. We illustrate this calibration with the 
two choices for $F_\Lambda$ as outlined in Sections~\ref{sse:discreteFtheta_weights_directalgorithm} 
and~\ref{sse:continuousFtheta_weights_directalgorithm}.

\subsubsection{Discrete
$F_\Lambda$}\label{sse:Optim_discrete_optimalthetadistribution_directalgorithm}

In the discrete case, we obtain
\begin{align}\label{eq:linearconditiononTheta_direct}
K\sum_{k=1}^{n_\Lambda} \frac{\bfone\{x_k\leq t\}p_k}{1-x_k}
 \approx\Psi(t\bfone), \quad t\in[0,1).
\end{align}
We propose to determine the $p_k$'s by enforcing
equality to hold in~\eqref{eq:linearconditiononTheta_direct} only for
$t=x_1,\dots,x_{n_\Lambda}$ which leads to the triangular system
\begin{align*}
 K\sum_{l=1}^k \frac{1}{1-x_l}p_l =\Psi(x_k\bfone), \quad k=1,\dots,n_\Lambda.
\end{align*}
Choosing the $x_k$'s as in the rejection sampling approach, we can solve for
the $p_k$'s using the following algorithm:
\begin{algorithm}\label{alg:ThetaDistribution} $\,$
\begin{enumerate}
\item Choose $n_\Lambda\in\NN$;
\item Define $x_k=1-(1/2)^{k-1}$, $k=1,\dots,n_\Lambda$;
\item Define $\widetilde{p}_1=\Psi(0,\dots,0)$ and $\widetilde{p}_k = \left(\Psi(x_k\bfone)-\Psi(x_{k-1}\bfone)\right)(1-x_k)$, for $k=2,\dots,n_\Lambda$;
\item Define $p_k=\widetilde{p}_k/(\sum_l\widetilde{p}_l)$.
\end{enumerate}
\end{algorithm}

\subsubsection{Continuous
$F_\Lambda$}\label{sse:Optim_continuous_optimalthetadistribution_directalgorithm}

In the continuous case, the optimization unfortunately cannot be done as easily
and explicitly as for the discrete case. In this case, the optimization on $K\in\RR$, $\gamma\in(0,1)$ and
$\beta>1$ is performed such that
\begin{align}\label{eq:continuous_lambda_tosolve_directalgorithm}
K\left[1+\frac{\gamma\left(1-\beta(1-t)^{\beta-1}\right)}{\beta-1}\right]
\approx \Psi(t\mathbf{1}), \quad t\in[0,1].
\end{align}

\section{Rare event analysis}\label{se:rare_event_analysis}

As the importance sampling technique is intended to be used in cases
where the functional $\Psi$ is large on sets which relate to rare events, we may
want to study the efficiency of the algorithm in a rare event setting. We shall
consider $\Psi^{(s)}(\textbf{u})$ as a functional that will take non-zero values only 
on a small probability set. Let $p^{(s)}=\EE\left[\Psi^{(s)}(\textbf{U})\right]$
be the probability of interest.
The rare event setting assumes that $\lim_{s\rightarrow 1} p^{(s)}
= 0$. For each $s$, we would select a new mixing distribution $F_\Lambda^{(s)}$, 
therefore changing the calibration of the proposal distribution
$F^{(s)}_{\mathbf{V}}$ and its sampling cost that we shall denote
$T(s)$. In the direct sampling algorithm, see
Section~\ref{se:DirectAlgorithm}, this sampling cost is finite and constant in
$s$, whereas it is of order $\EE[(1-\Lambda)^{-1}]$, see Theorem
\ref{thm:expectationNV}, in the rejection sampling algorithm from
Section~\ref{se:RejectionAlgorithm}.

Let $\muIS^{(s)} = n^{-1} \sum_{i=1}^n \Psi^{(s)}(\bfV_i)
w^{(s)}(\bfV_i)$ be the importance sampling estimate. In a rare-event setting,
we would ideally aim for a bounded relative error as $s\rightarrow1$, see
Chapter VI in \cite{AsmussenGlynn2007}, that is 
\begin{align}\label{eq:bounded_rel_error}
\limsup\limits_{s\rightarrow1}\frac{\var\left[\muIS^{(s)}\right]}{\left(p^{(s)}\right)^2}T(s)<\infty.
\end{align}

Replacing $\var\left[\muIS^{(s)}\right]$ by its upper bound
$n^{-1}\EE\left[\Psi^{(s)}(\bfV)^2 w^{(s)}(\bfV)^2\right]$, we shall aim for an
algorithm that satisfies
\begin{align}\label{eq:bounded_rel_error_2}
\limsup\limits_{s\rightarrow1}\frac{n^{-1}\EE\left[\Psi^{(s)}(\bfV)^2 w^{(s)}(\bfV)^2\right]}{\left(p^{(s)}\right)^2}T(s)<\infty.
\end{align}
Note first that the optimality condition~\eqref{eq:optimalcondition} guarantees
that $\EE\left[\Psi^{(s)}(\bfV)^2 w^{(s)}(\bfV)^2\right]/\left(p^{(s)}\right)^2\propto
1$. We now assume a mild condition for the calibration of $F_\Lambda^{(s)}$ that
will be needed to obtain the efficiency criteria~\eqref{eq:bounded_rel_error_2}.

\begin{condition}\label{cond:rareevent}
For all $s > 0$, the discrete distribution of $\Lambda$ is constructed such
that there exists $k\in\{1,\dots, n_\Lambda\}$ with $x_k= s$ and $p_k> 0$.
\end{condition}

We first study the case of rejection sampling from
Section~\ref{se:RejectionAlgorithm}, limiting ourselves to the setup
of a discrete distribution for $\Lambda$. Although typical rare event sets in
the literature consider the sum of margins, we will consider the maximum
instead, which allows us to stay within a rare event framework since $\{\max_i
u_i>s\}\subseteq\{\sum_{i}u_i > s\}$, $\bfu\in[0,1]^d$.

\begin{theorem}\label{thm:rareevent_rejection}
Assume that $\Psi^{(s)}(\textbf{u})=\bfone\{\max_i u_i>s\}$ and that the
proposal distribution $F_\bfV$ and the corresponding weight function $w(\bfu)$
are chosen as in Section~\ref{se:RejectionAlgorithm}. In addition, assume
$F_\Lambda$ is a discrete distribution with a finite number $n_\Lambda$ of atoms 
$\PP[\Lambda = x_k]=p_k, \quad k=1,\dots, n_\Lambda, \quad  0= x_1<\dots<
x_{n_\Lambda} < 1$, calibrated as in
Algorithm~\ref{alg:ThetaDistribution_rejection} and that the
Condition~\ref{cond:rareevent} holds.
Denote $k^*_\bfu=\max\{1\leq k\leq n_\Lambda:x_k\leq \max_i u_i\}, \bfu \in[0,1)^d$.
Then
\begin{align*}
\limsup\limits_{s\rightarrow1}\frac{\EE\left[\Psi^{(s)}(\bfV)^2 w^{(s)}(\bfV)^2\right]}{\left(p^{(s)}\right)^2}<\infty.
\end{align*}
\end{theorem}

\textit{Proof.}
Under Condition~\ref{cond:rareevent}, we have that
$x_{k^*_\bfu}\geq s$ on the event $\{\max_i u_i>s\}$. Therefore,
\begin{align}\label{eq:bounded_rel_error_rejection}
&\int_{[0,1]^d}\Psi^{(s)}(\bfv)^2
w^{(s)}(\bfv)^2\text{d}F_{\bfV}(\bfv)=\int_{[0,1]^d}\Psi^{(s)}(\bfu) w^{(s)}(\bfu)\text{d}C(\bfu)\nonumber\\
&=
\int_{[0,1]^d}\Psi^{(s)}(\bfu)\left(\sum_{k=1}^{n_\Lambda}\widetilde{p}_k\right)\left(\sum_{k=1}^{n_\Lambda}\frac{\bfone\{x_k\leq
\max_i u_i\}\widetilde{p}_k}{1-C(x_k\bfone)}\right)^{-1}\text{d}C(\bfu)\nonumber\\
&=
\int_{[0,1]^d}\Psi^{(s)}(\bfu)\left(\sum_{k=1}^{n_\Lambda}\Psi^{(s)}(x_k\bfone)(C(x_k\bfone)-C(x_{k-1}\bfone))\right)\Psi^{(s)}(x_{k^*_\bfu}\bfone)^{-1}\text{d}C(\bfu)\nonumber\\
&=
\int_{[0,1]^d}\Psi^{(s)}(\bfu)\left(C(x_{n_\Lambda}\bfone)-C(s\bfone)\right)\text{d}C(\bfu)\nonumber\\
&\leq
\left(1-C(s\bfone)\right)\int_{[0,1]^d}\Psi^{(s)}(\bfu)\text{d}C(\bfu)= (p^{(s)})^2,
\end{align}
which proves the theorem. \qed

Note that Theorem~\ref{thm:rareevent_rejection} guarantees a bounded relative
error as in~\eqref{eq:bounded_rel_error} whenever $\limsup_{s\rightarrow1} T(s)<\infty.$ This would not hold for the
rejection algorithm. Indeed, since
$\EE[(1-\Lambda)^{-1}]=\sum_{k=1}^{n_\Lambda}p_k /(1-x_k)$, we obtain in virtue of 
Theorem~\ref{thm:expectationNV} that $\limsup_{s\rightarrow1}T^{(s)}=\infty$
under Condition~\ref{cond:rareevent}.

In the case of direct sampling, we can prove the corresponding version of
Theorem~\ref{thm:rareevent_rejection} by taking 
$\Psi^{(s)}(\textbf{u})=\bfone\{u_i>s \}$ for any $i=1,\dots,d$ and 
$k^*_\bfu=\max\{1\leq k\leq n_\Lambda:x_k\leq u_i\}, \bfu \in[0,1)^d$. 
Since this algorithm has a computational cost $T(s)$ constant in $s$, it shall then be 
prefered to the rejection sampling algorithm in rare event settings, although it
may require more implementation efforts. 
 
The calibration of the proposal distribution $F_{\bfV}$ is profiled on the assumption that $\Psi(\textbf{u})\approx\Psi(\max_i u_i \bfone)$. In Theorem~\ref{thm:rareevent_rejection} we have been able to show that when $\Psi(\textbf{u})=\bfone\{\max_i u_i > s\}$ for some $s\in(0,1)$, i.e. when the assumption holds with equality, then $\EE[\Psi(\bfV)^2 w(\bfV)^2]\leq \EE[\Psi(\bfU)]^2$. By Jensen's inequality we obtain that $\EE[\Psi(\bfV)^2 w(\bfV)^2]\leq \EE[\Psi(\bfU)^2]$, and thus that $\var(\widehat{\mu}_n)\leq \var(\mu_n)$, so a smaller estimator's variance. Although the assumption that $\Psi(\textbf{u})\approx\Psi(\max_i u_i \bfone)$ is typical of application in insurance mathematics, it does not often hold with equality and thus cannot be easily incorporated into an analytical framework that would allow to prove a certain variance reduction factor. However, we illustrate in the numerical Case Study of Section~\ref{se:casestudy} that we obtain a substantial variance reduction for several typical insurance problems. 
 
\section{Case study}\label{se:casestudy}

In this section, we illustrate the performance of our importance sampling
algorithms for functionals $\Psi$ relevant for insurance applications. We
shall use the two importance sampling algorithms defined in
Section~\ref{se:RejectionAlgorithm} and Section~\ref{se:DirectAlgorithm} on the
same example. We use three random vectors, of dimension $d=2$, $d=5$, and
$d=25$, respectively. Our case study will assume that marginal distributions of
$\bfX=(X_1,\dots,X_d)$ are lognormal, parametrized as $X_j\sim \text{LN}(10-0.1j,1+0.2j)$,
$j=1,\dots,d$, which yields equal expectation for each margin, i.e,
$\EE[X_j]=36\,315.5$ and standard deviation $\EE[X_j]\sqrt{e^{1+0.2j}-1}$.
We will consider two examples of copulas, namely Clayton and Gumbel. Kendall's tau, 
see, e.g., Section 5.1.1 in~\cite{nelsen},  between each pair of margins is
$1/3$, which yields a Clayton parameter of~1 and a Gumbel parameter of~1.5. Note
that our importance sampling method does not rely on particular assumptions on
the copula. In consequence, the general behavior of the algorithm does not
significantly change with the strength of the dependence. This case study has
been implemented using the \textsf{R} package \texttt{copula}.

We investigate the estimation of the following five functionals of $\bfX$. All are formulated in 
terms of the aggregate losses $S=\sum_{j=1}^{d}X_j$, which is inspired by risk
aggregation problems arising frequently in actuarial practice:
\begin{itemize}
\item $\EE[\max\{S-T,0\}]$, which is the fair premium of a stop-loss cover  with
deductible $T$. For $T$ we use $T=10^5 d$, which is approximately $3$ times the
expectation of $S$;
\item $\VaR_{0.995}(S)$ and $\ES_{0.99}(S)$, which are the risk measures
determining solvency capital under Solvency~II and the Swiss Solvency Test,
respectively (see \cite{ssttechspec} and \cite{ceiopsQIS5});
\item $\EE[X_1\,|\,S>F^{-1}_S(0.99)]$ and $\EE[X_d\,|\,S>F^{-1}_S(0.99)]$, which
represent the capital allocated to the first and last margin under the Euler 
principle, see \cite{Tasche2008}.
\end{itemize}

For ease of calibration and simulation, we use the discrete framework for $F_\Lambda$. 
As we want to use the same sample  to estimate all objective functions, we only
conduct one calibration of $F_\Lambda$ for each problem dimension. Recall from
Section~\ref{se:Motivation} that $\VaR_\alpha$ and $\ES_\alpha$ cannot be written as 
an expectation of type $\EE[\Psi_0(\bfX)]$. We thus calibrate $F_\Lambda$ using
the stop-loss objective function $\Psitilde(\bfu) = \max\{ \sum_{j=1}^{d} F_j^{-1}(u_j)-T,0\}$. 
This is non-zero only for  $\sum_{j=1}^{d} F_j^{-1}(u_j)$ above the deductible
$T$, so that calibration with this function will favour a high concentration of 
distorted samples in the region of interest for our applications. Note that the
calibration of $F_\Lambda$ depends on the choice of copula and of the importance 
sampling algorithm, see
Sections~\ref{sse:Optim_discrete_optimalthetadistribution_rejectionalgorithm} 
and~\ref{sse:Optim_discrete_optimalthetadistribution_directalgorithm}. The
number of discretization points is set to $n_\Lambda=10$. As shown in Table~\ref{ta:weights}, 
the highest point $x_{10}=1-(1/2)^9\approx0.998$, which is well beyond the highest $\VaR$ level under consideration. 
In order to satisfy Condition~\ref{cond:Lambdamass0}, we manually set 
the weight of $x_0$ to be $p_0=0.1$ and decrease the other weights
proportionally. The weights $p_k$ for $k=1,\dots,n_\Lambda$ resulting from the
calibration using the Gumbel copula and the rejection sampling approach are
shown in Table~\ref{ta:weights} for dimensions $d=2, 5\text{ and } 25.$

\begin{table}[ht]
{\centering
\begin{tabular}{cccccccccccc}\toprule
& $k$ & 1 & 2 & 3 & 4 & 5 & 6 & 7 & 8 & 9 & 10 \\
\cmidrule(lr){2-2}\cmidrule(lr){3-12}
& $x_k$ & 0.000 & 0.500 & 0.750 & 0.875 & 0.937 & 0.968 & 0.984 & 0.992 & 0.996 & 0.998 \\ 
$d=2$ & $p_k$ & 0.100 & 0.000 & 0.000 & 0.000 & 0.115 & 0.325 & 0.206 & 0.128 & 0.787 & 0.048 \\
$d=5$ & $p_k$ & 0.100 & 0.000 & 0.000 & 0.000 & 0.129 & 0.302 & 0.202 & 0.131 & 0.084 & 0.053 \\
$\phantom{0}d=25$ & $p_k$ & 0.100 & 0.000 & 0.000 & 0.000 & 0.022 & 0.252 &
0.216 & 0.174 & 0.135 & 0.102 \\
\bottomrule
\end{tabular}
\caption{Calibrated probability weights
$p_k$ using the Gumbel copula.}\label{ta:weights} }
\end{table}

\begin{figure}[ht]
{\centering
\scalebox{1}{\includegraphics[width = 7.3cm, height = 7cm]{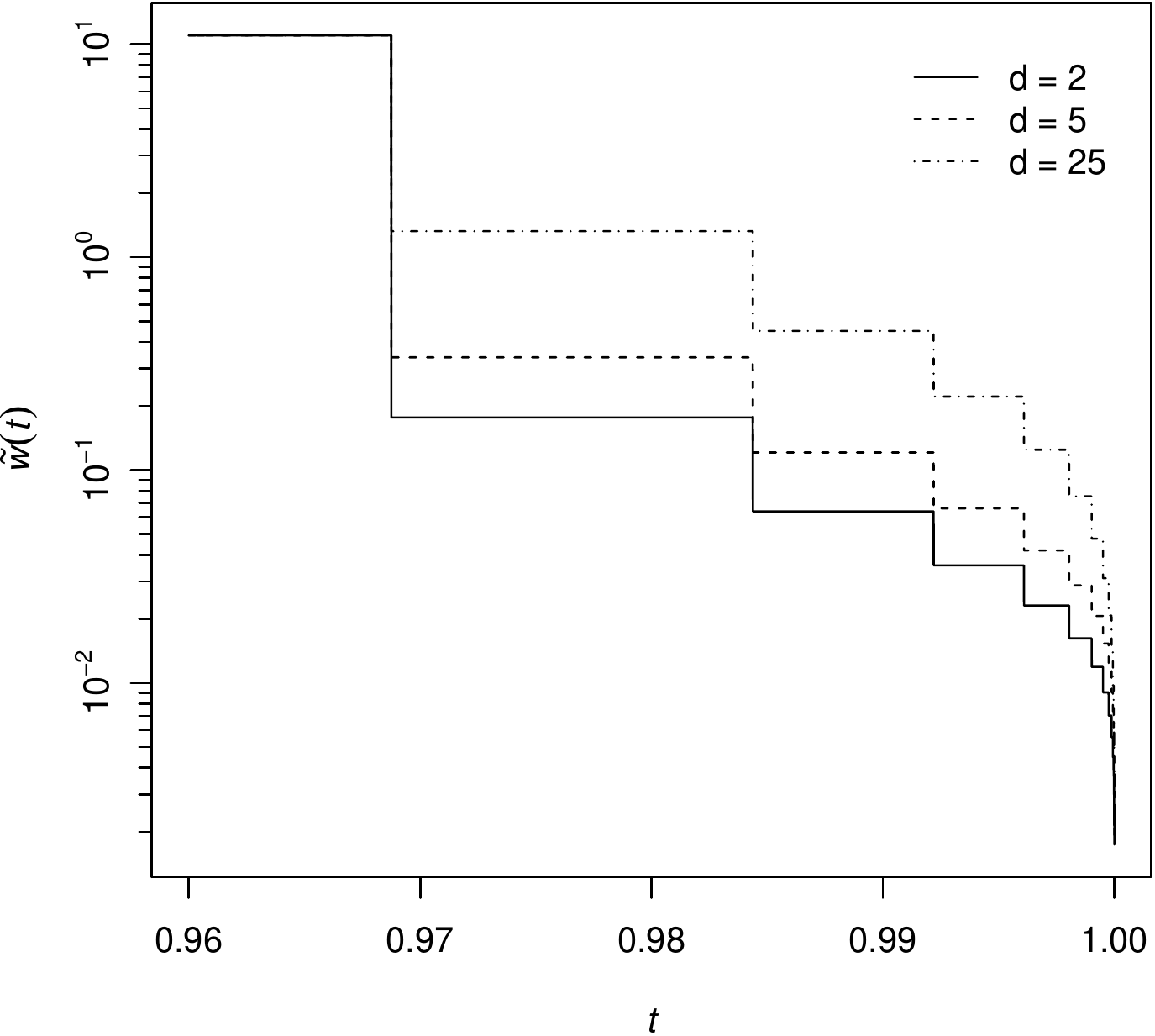}}
\hspace*{0.5cm}
\scalebox{01}{\includegraphics[width =7.3cm, height = 7cm]{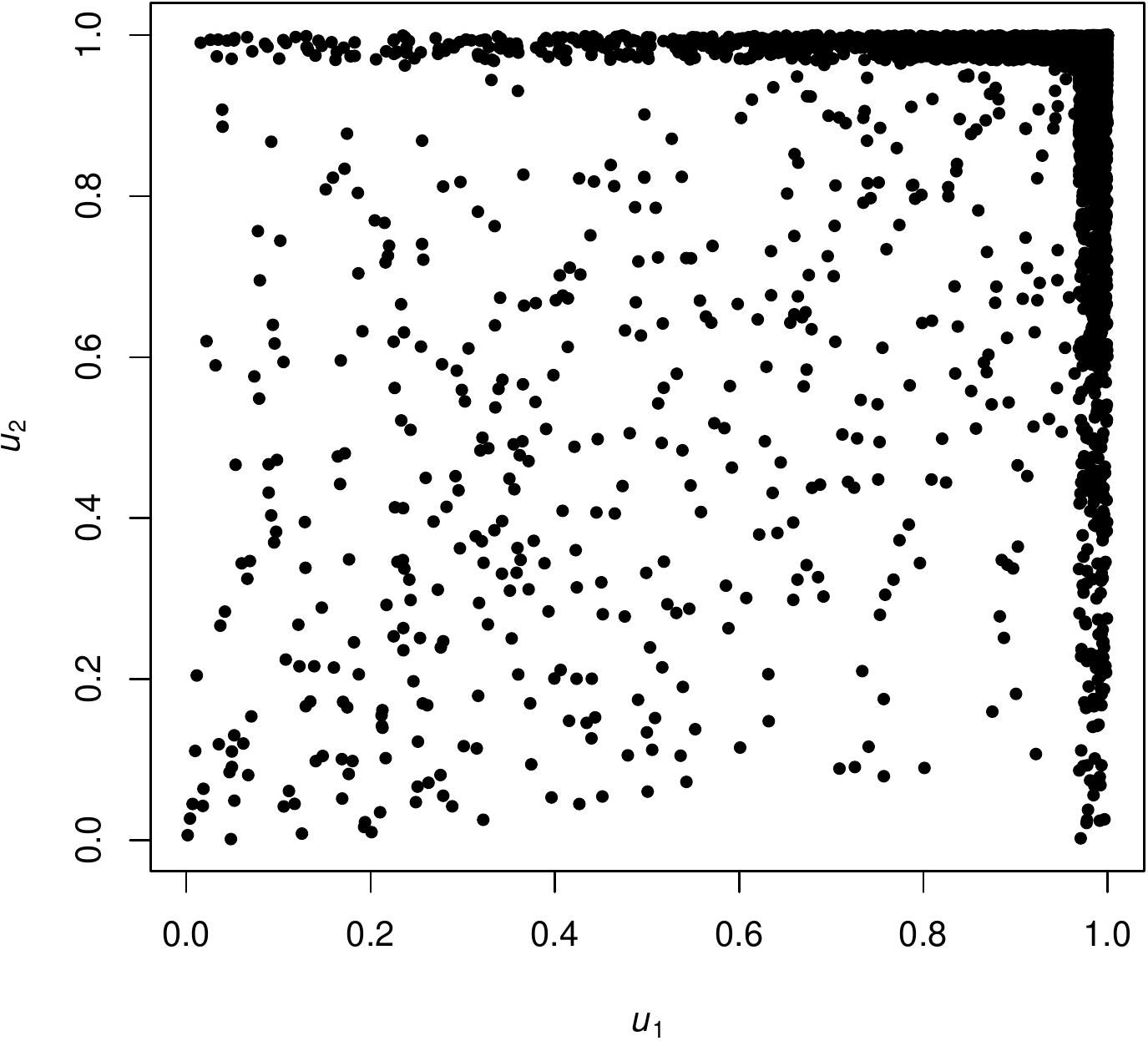}}
\caption{
\textit{Left:} Discrete weight function $\wtilde(t)=w(t\bfone)$ for $t\in[0.96,1]$ for
the Gumbel copula, for dimensions $d\in\{2, 5, 25\}$ and using the rejection
sampling approach. From $0$ to $0.968$ the function $\wtilde$ is constant.
\textit{Right:} A scatter plot of 5\,000 samples of $\bfV$ sampled using the
rejection sampling approach.
}\label{fig:weightfunction} }
\end{figure}

The weight functions $\wtilde(\cdot)=w(\cdot\bfone)$ and a scatter plot of 5\,000
samples of $\bfV$ are plotted in Figure~\ref{fig:weightfunction}, when the reference
copula is Gumbel and the rejection sampling approach is used. Given the setup
of this case study, it is easy to check that Lemma~\ref{thm:consistency} is satisfied.
Due to the construction of $F_\bfV$, more samples lie close to the upper or
right border than what would be observed for a copula sample.

As the objective functions use estimates of the distribution function of $S$, we
normalize the sample weights to sum to $1$. This further reduces the estimation 
error as advocated in Section 4.2.2\ in \cite{Geweke2005} or Section 2.5.2 in 
\cite{Liu2008}.

In order to assess the improvements provided by importance sampling, we present
a simulation study for $d\in\{2, 5, 25\}$. We use a sample size of $n=10\,000$
to calculate the importance sampling estimators $\muIS$ for each of
the two algorithms and the standard Monte Carlo estimators $\muMC$ for all
objective functions. A total of 500 repetitions is used to obtain an empirical
distribution of these estimators, and thus to estimate their variance.
Although the sampling size has an impact on the value of the sample variance, it
should not play a significant role in the study of algorithms efficiency. The
results are presented in Tables~\ref{ta:casestudyresults_Gumbel_rejection},
\ref{ta:casestudyresults_Clayton_rejection} for the Gumbel and Clayton cases
using the rejection algorithm, and Tables~\ref{ta:casestudyresults_Gumbel_direct}, 
\ref{ta:casestudyresults_Clayton_direct} for the Gumbel and Clayton cases using
the direct algorithm. Although the value of the estimates may be different 
depending on which algorithm is used for sampling, we only present the reference
value obtained from the plain Monte Carlo simulation since our empirical study 
has shown only negligible differences. The main results in these tables are in 
the form of variance reduction factors, which represent the sample variance of 
the plain Monte Carlo estimator divided by the sample variance of our importance
sampling estimator. 

\begin{table}[h]
{\centering
\begin{tabular}{lrrrrrr}\toprule
& \multicolumn{2}{c}{$d=2$}& \multicolumn{2}{c}{$d=5$}& \multicolumn{2}{c}{$d=25$}\\
\cmidrule(lr){2-3}\cmidrule(lr){4-5}\cmidrule(lr){6-7}
Objective function & Ref. val. & Red. fact. &
Ref. val. & Red. fact. & Ref.
val. & Red. fact. \\
\cmidrule(lr){1-7}
$\EE[(S-T)^+]$             &   10\,498 & 80.8 &    29\,648 & 39.1 &    310\,499
& 17.3  \\
$\VaR_{0.995}(S)$          & 645\,162 & 12.4 & 1\,795\,071 & 11.5 & 15\,183\,823
& 9.9  \\
$\ES_{0.99}(S)$            & 774\,616 & 18.6 & 2\,241\,589 & 17.5 & 24\,541\,482
& 13.9  \\
$\EE[X_1|S>F^{-1}_S(0.99)]$& 351\,077 & 21.4 &   332\,560 & 19.3 &    324\,231 &
11.3  \\
$\EE[X_d|S>F^{-1}_S(0.99)]$& 423\,539 & 22.3 &   570\,105 & 18.1 &  1\,676\,897
& 17.7 \\\bottomrule
\end{tabular}
\caption{Reference values (Ref. val.) of the objective functions and variance
reduction factors (Red. fact.) with $F_\lambda$ discrete and $C$ Gumbel
copula, using the rejection sampling
algorithm.}\label{ta:casestudyresults_Gumbel_rejection} }
\end{table}

\begin{table}[h]
{\centering
\begin{tabular}{lrrrrrr}\toprule
& \multicolumn{2}{c}{$d=2$}& \multicolumn{2}{c}{$d=5$}& \multicolumn{2}{c}{$d=25$}\\
\cmidrule(lr){2-3}\cmidrule(lr){4-5}\cmidrule(lr){6-7}
Objective function & Ref. val. & Red. fact. & Ref. val. & Red. fact. & Ref. val.
& Red. fact. \\
\cmidrule(lr){1-7}
$\EE[(S-T)^+]$             &   7\,765 & 63.08 &    13\,657 & 23.59 &    119\,531
& 9.05  \\
$\VaR_{0.995}(S)$          & 526\,254 & 14.14 & 1\,101\,395 & 10.60 &
7\,235\,669 & 6.05  \\
$\ES_{0.99}(S)$            & 610\,928 & 19.74 & 1\,272\,925 & 14.84 &
9\,963\,262 & 8.47  \\
$\EE[X_1|S>F^{-1}_S(0.99)]$& 259\,814 & 31.18 &   139\,127 & 19.18 &    68\,702
& 5.62  \\
$\EE[X_d|S>F^{-1}_S(0.99)]$& 351\,113 & 26.04 &   384\,475 & 16.92 & 
1\,009\,675 & 15.81 \\\bottomrule
\end{tabular}
\caption{Reference values (Ref. val.) of the objective functions and variance
reduction factors (Red. fact.) with $F_\lambda$ discrete and $C$ Clayton
copula, using the rejection sampling
algorithm.}\label{ta:casestudyresults_Clayton_rejection} }
\end{table}

\begin{table}[h]
{\centering
\begin{tabular}{lrrrrrr}\toprule
& \multicolumn{2}{c}{$d=2$}& \multicolumn{2}{c}{$d=5$}& \multicolumn{2}{c}{$d=25$}\\
\cmidrule(lr){2-3}\cmidrule(lr){4-5}\cmidrule(lr){6-7}
Objective function & Ref. val. & Red. fact. & Ref. val. & Red. fact. & Ref. val.
& Red. fact. \\
\cmidrule(lr){1-7}
$\EE[(S-T)^+]$             &   10\,498 & 116.03 &    29\,648 &  80.27 &   
310\,499 & 21.71  \\
$\VaR_{0.995}(S)$          & 645\,162 & 14.25 & 1\,795\,071 & 15.83 & 15\,183\,823
& 8.97  \\
$\ES_{0.99}(S)$            & 774\,616 & 20.98 & 2\,241\,589 & 19.78 &
24\,541\,482 & 12.14  \\
$\EE[X_1|S>F^{-1}_S(0.99)]$& 351\,077 & 23.84 &   332\,560 & 19.01 &    324\,231
& 11.85  \\
$\EE[X_d|S>F^{-1}_S(0.99)]$& 423\,539 & 23.87 &   570\,105 & 20.67 & 
1\,676\,897 & 19.52 \\\bottomrule
\end{tabular}
\caption{Reference values (Ref. val.) of the objective functions and variance
reduction factors (Red. fact.) with $F_\lambda$ discrete and $C$ Gumbel
copula, using the direct sampling
algorithm.}\label{ta:casestudyresults_Gumbel_direct} }
\end{table}

\begin{table}[h]
{\centering
\begin{tabular}{lrrrrrr}\toprule
& \multicolumn{2}{c}{$d=2$}& \multicolumn{2}{c}{$d=5$}& \multicolumn{2}{c}{$d=25$}\\
\cmidrule(lr){2-3}\cmidrule(lr){4-5}\cmidrule(lr){6-7}
Objective function & Ref. val. & Red. fact. & Ref. val. & Red. fact. & Ref. val.
& Red. fact. \\
\cmidrule(lr){1-7}
$\EE[(S-T)^+]$             &   7\,765 & 72.17 &    13\,657 & 22.34 &    119\,531
& 5.82  \\
$\VaR_{0.995}(S)$          & 526\,254 & 14.74 & 1\,101\,395 & 11.05 &
7\,235\,669 & 6.33  \\
$\ES_{0.99}(S)$            & 610\,928 & 20.18 & 1\,272\,925 & 12.60 &
9\,963\,262 & 5.23  \\
$\EE[X_1|S>F^{-1}_S(0.99)]$& 259\,814 & 31.41 &   139\,127 & 14.93 &    68\,702
& 10.55  \\
$\EE[X_d|S>F^{-1}_S(0.99)]$& 351\,113 & 25.57 &   384\,475 & 14.84 & 
1\,009\,675 & 10.98 \\\bottomrule
\end{tabular}
\caption{Reference values (Ref. val.) of the objective functions and variance
reduction factors (Red. fact.) with $F_\lambda$ discrete and $C$ Clayton
copula, using the direct sampling
algorithm.}\label{ta:casestudyresults_Clayton_direct} }
\end{table}

\begin{figure}[h]
{\centering
\scalebox{1.6}{\includegraphics[width
=7.3cm]{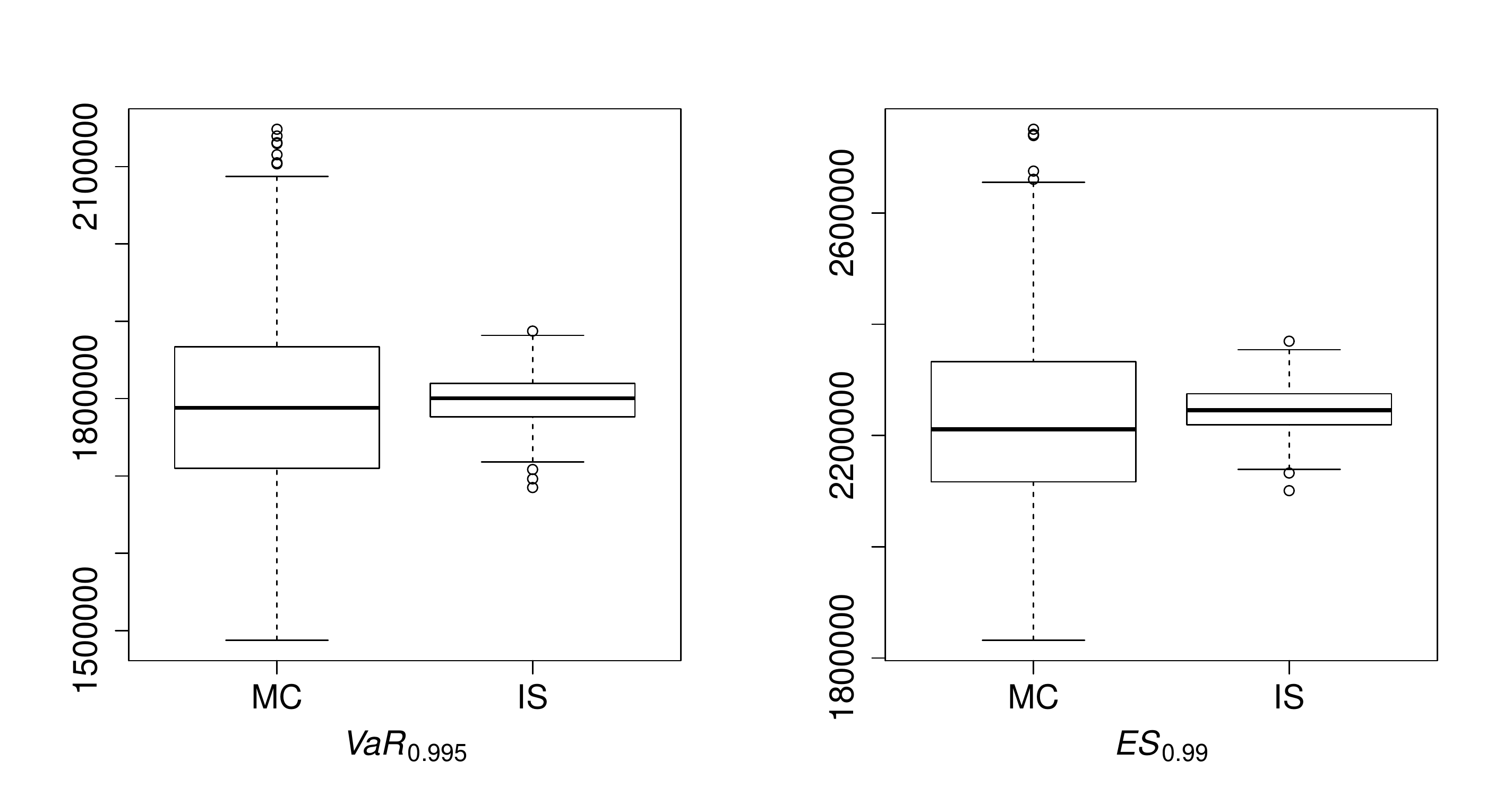}} \\
\scalebox{1.6}{\includegraphics[width
=7.3cm]{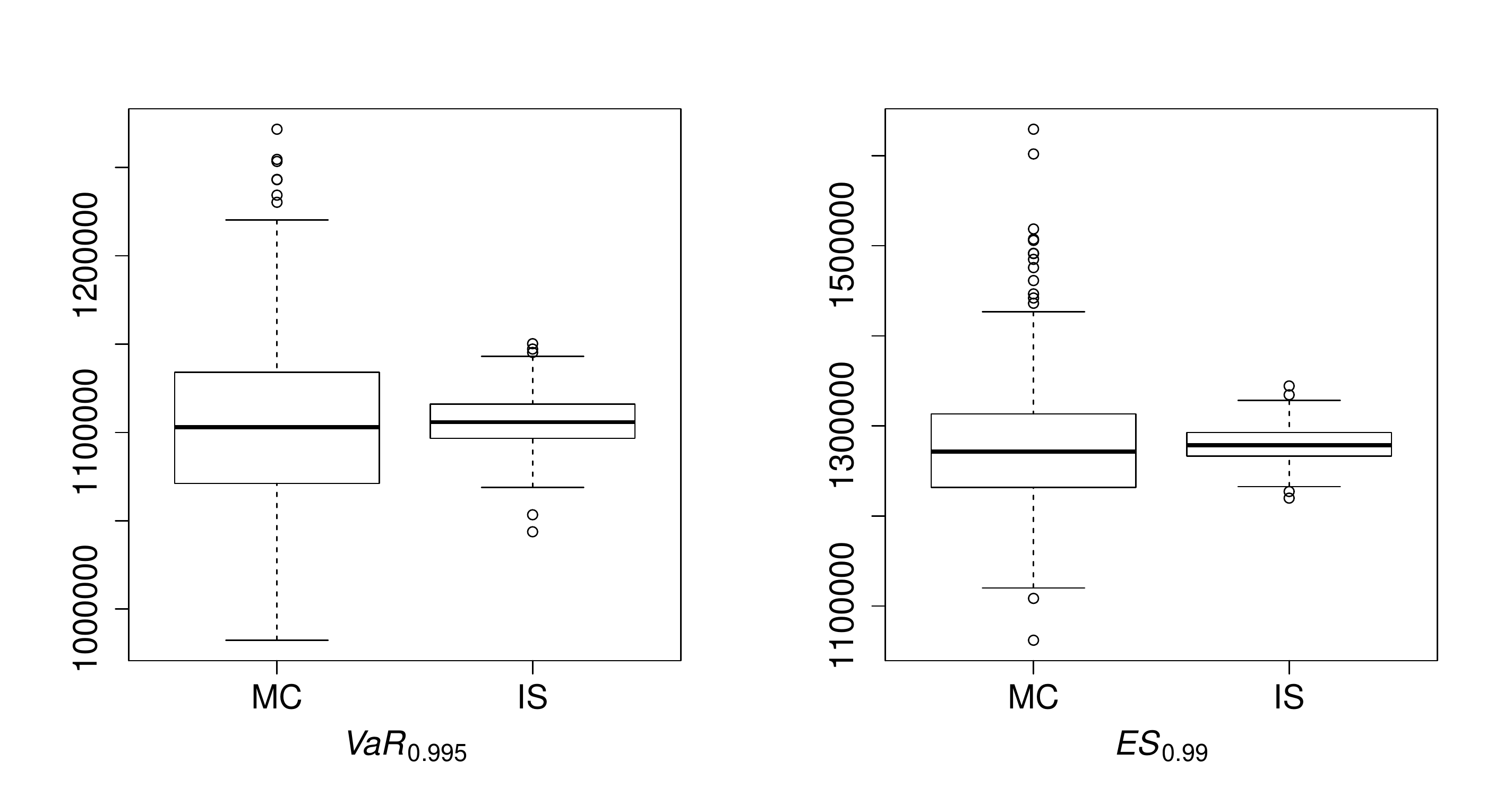}}

\caption{Boxplots for the importance sampling estimators of
$\VaR_{0.995}(S)$ and $\ES_{0.99}(S)$ from the $N = 500$ independent copies of
the estimator, using the rejection sampling algorithm in the Gumbel
(\textit{top}) and Clayton (\textit{bottom}) cases for $d=5$.
}
\label{fig:boxplot_RM_rejection} }
\end{figure}

\begin{figure}[h]
{\centering
\scalebox{1.6}{\includegraphics[width
=7.3cm]{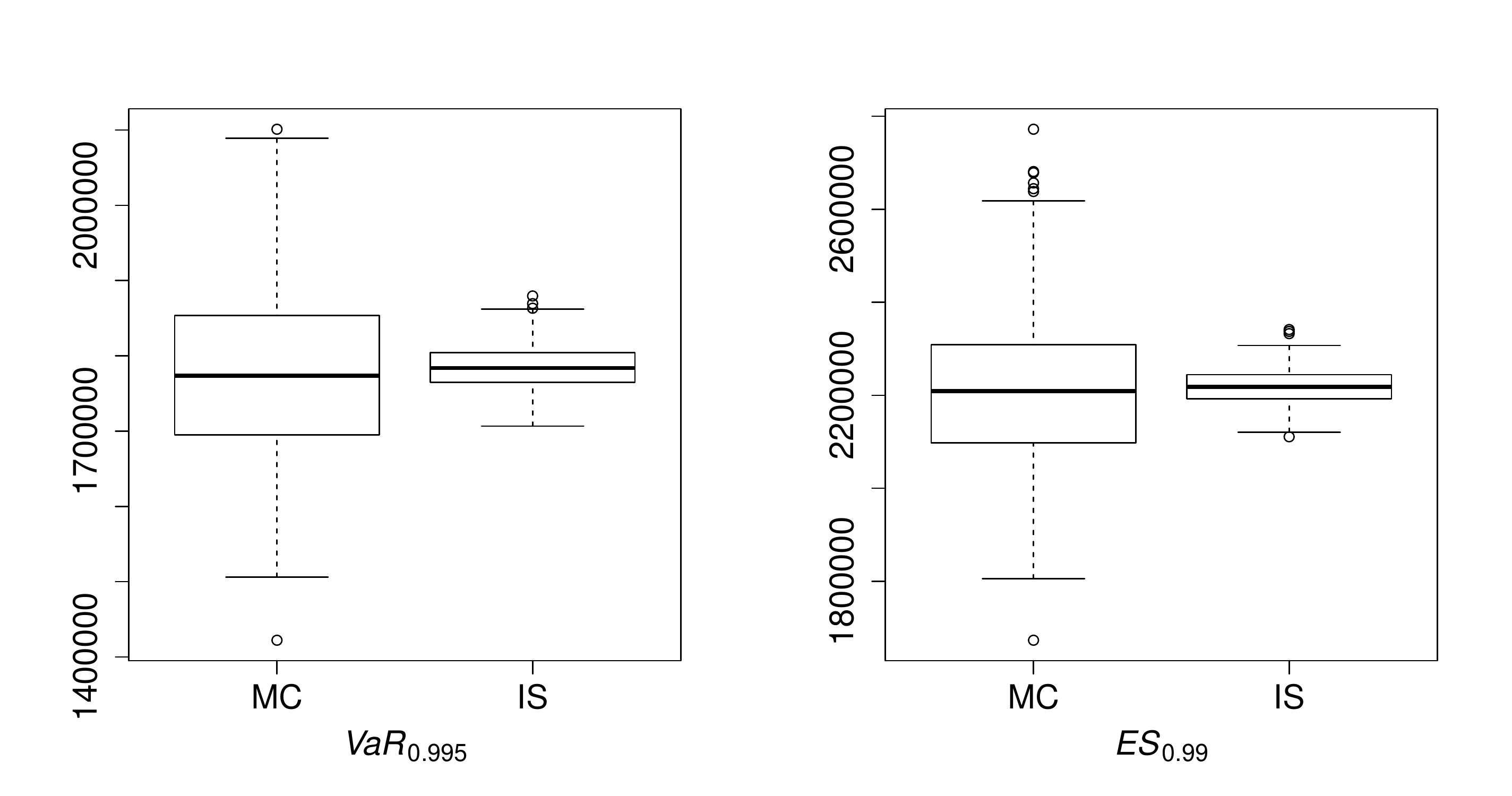}}\\
\scalebox{1.6}{\includegraphics[width
=7.3cm]{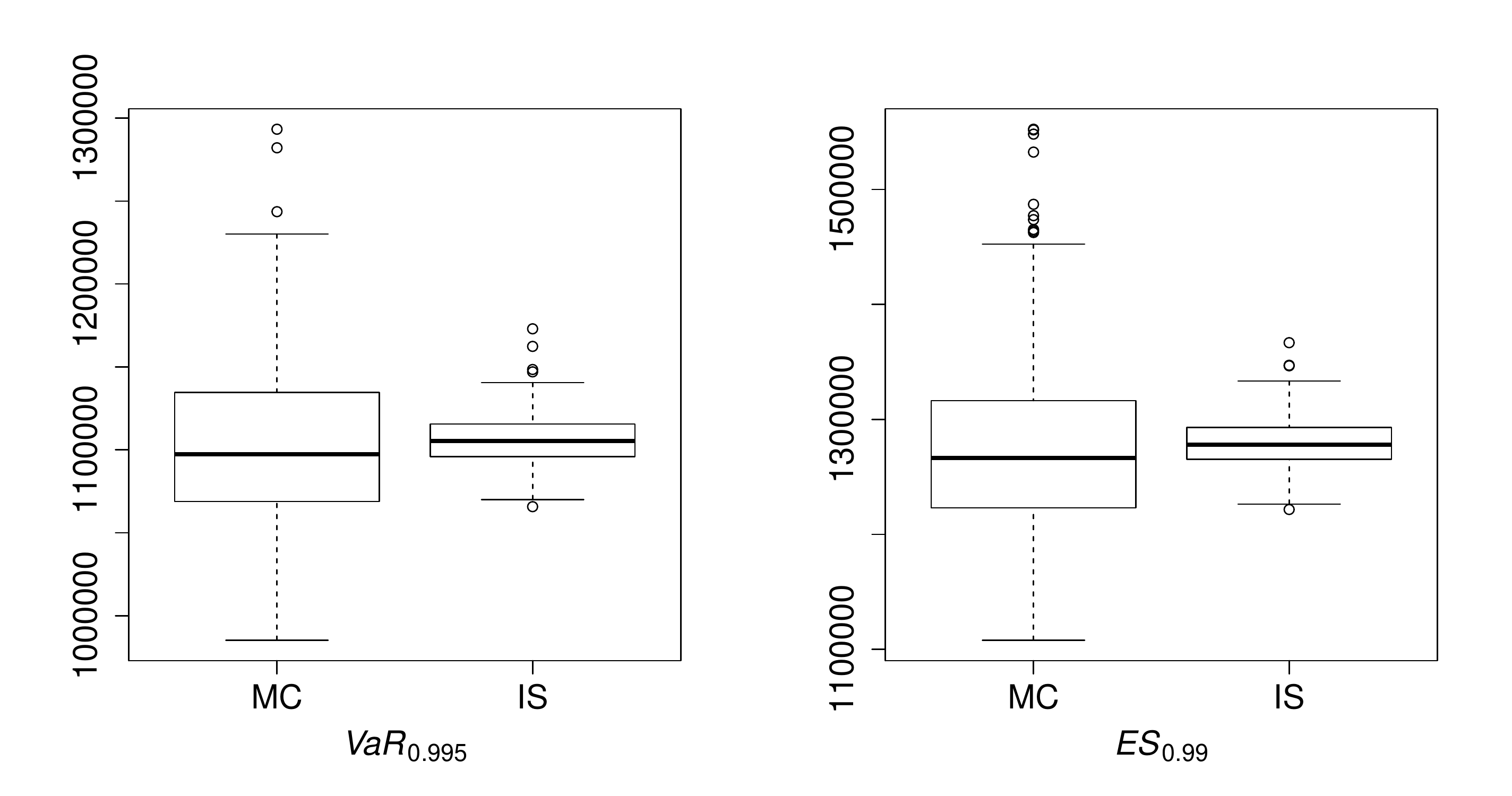}}

\caption{Boxplots for the importance sampling estimators of
$\VaR_{0.995}(S)$ and $\ES_{0.99}(S)$ from the $N = 500$ independent copies of
the estimator, using the direct sampling algorithm in the Gumbel (\textit{top})
and Clayton (\textit{bottom}) cases for $d=5$.
}
\label{fig:boxplot_RM_direct} }
\end{figure}

Tables~\ref{ta:casestudyresults_Gumbel_rejection},
\ref{ta:casestudyresults_Clayton_rejection},
\ref{ta:casestudyresults_Gumbel_direct}
and~\ref{ta:casestudyresults_Clayton_direct} show that the importance sampling 
algorithms greatly decrease the estimation error for all objective functions. 
It is not surprising that the largest reduction is for the stop-loss cover,
since $F_\Lambda$ is calibrated to this functional.
A larger reduction for the other functionals could be achieved with a specific 
calibration for each of them. The smallest reduction factors are for $\VaR_{0.995}(S)$, 
because this functional does not depend on the tail behaviour of $S$ beyond the
$99.5\%$ quantile, where the largest gain in accuracy is obtained by our
importance sampling approach. The variance reduction can also be observed from
the boxplots in Figures~\ref{fig:boxplot_RM_rejection}
and~\ref{fig:boxplot_RM_direct} that allow us to compare the entire distribution 
of the plain Monte Carlo to the importance sampling
estimators for $\VaR_{0.995}(S)$ and $\ES_{0.99}(S)$, using the two algorithms
for $d=5$. We note that the bias indeed appears negligible and that variance of
these empirical distributions has been greatly reduced by the importance sampling
approaches.

In order to fairly assess the efficiency of the method and to compare the two
sampling algorithms, one should divide the variance reduction factors in
Tables~\ref{ta:casestudyresults_Gumbel_rejection}
and~\ref{ta:casestudyresults_Clayton_rejection} by the expected waiting time of
the rejection sampling algorithm, $\EE[N_{\bfV}]$, given in Table~\ref{ta:expected_waiting}.

\begin{table}[h]
{\centering
\begin{tabular}{lrrr}\toprule
& $d=2$ & $d=5$ & $d=25$ \\
\cmidrule(lr){1-4}
Clayton $(\theta = 1)$ & 44.16 & 19.48 & 6.89 \\
Gumbel $(\theta = 1.5)$ & 54.69 & 31.11 & 15.83 \\\bottomrule
\end{tabular}
\caption{Expected waiting time, $\EE[N_{\bfV}]$, of the
rejection sampling algorithm.}\label{ta:expected_waiting} }
\end{table}

In most cases, the expected waiting time is larger
than the variance reduction ratio, hence rendering the rejection algorithm
inefficient for this case study. We recall that this waiting time issue is only
a concern when using the rejection sampling algorithm from Section~\ref{se:RejectionAlgorithm}.

Although the conditional sampling algorithm might be a
bit more computationally intensive than the direct sampling of the copula $C$,
e.g., when inverting the conditional distributions in Algorithm~\ref{alg:conditionalcopulaSampling} for certain
copulas, this complexity is insignificant and does not become more pronounced if
one puts more mass of $\Lambda$ towards 1. For this reason, we conclude that the
efficiency of the importance sampling method is not reduced with the conditional
sampling approach. 

\section{Conclusion}\label{se:conclusion}

We proposed an importance sampling approach for copula models with two
algorithms, specifically designed for problems arising
frequently in insurance and financial applications.

The starting point for the construction of an alternative sampling distribution
was to consider the copula conditional on the event that some of its
components exceed a certain threshold. In the rejection sampling approach, 
we require that the maximum of all components exceeds the threshold. In the
direct sampling approach, we require that a specific component exceeds the
threshold. The proposal distribution has then  been constructed by mixing over
different thresholds.

In order to minimize the estimation error of the importance sampling estimator,
we proposed several procedures to set up and optimize the mixing distribution. 
Unlike other importance sampling approaches, our method does not have
requirements on the original copula and it can be applied to any copula from
which sampling is feasible.

The variance reduction of our approach has only been shown analytically for a simplified case. Through a case study inspired by a typical insurance application, we have shown
that the rejection and the direct sampling algorithms are able to largely reduce
simulation errors in more general estimation problems relevant to actuarial practitioners.

In the rejection sampling approach, sampling the proposal distribution can
easily be implemented through a rejection algorithm, which only requires that samples from the original copula can be drawn. 
It is acknowledged that the computational cost of the algorithm is increased 
due to the rejection sampling procedure, which however is not always a
disadvantage.
In addition, the direct sampling algorithm based on the inversion of
conditional distributions has been proposed. Although it requires a more advanced implementation, 
this algorithm has the striking advantage that it has a reduced
computational complexity, of order of the cost of sampling $C$ and that it does
not depend on the calibration of the proposal distribution $F_\bfV$. We have
also shown that the later algorithm is efficient in a rare-event setting in the
sense of \cite{AsmussenGlynn2007}.

For further research, we emphasize the problem of sampling from conditional
distributions such as the $C^{[\lambda]}$'s proposed in
Section~\ref{se:RejectionAlgorithm} and \ref{se:DirectAlgorithm}. One could aim
at finding copulas for which the sampling of $C^{[\lambda]}$ is sufficiently
simple, an example is given in
Appendix~\ref{app:AnalyticCtheta}. More generally, families of copulas such that
$C^{[\lambda]}$ stays within the same class for all $\lambda\in[0,1]$ could be
of interest. Note that copulas that are invariant under conditioning on
subregions of $[0,1]^d$ have been investigated in
\cite{CharpentierTruncInvariant2006}, \cite{JavidTruncInvariant2009} or
\cite{DuranteTruncInvariant2012}. However, the conditional regions are always $d$-rectangles such as $\prod_{i=1}^d[\alpha_i,\beta_i]\subseteq[0,1]^d.$ The regions we condition on in the rejection sampling approach, $\left\{\max_i
u_i>\lambda\right\},$ are unions of stripes. 

\section*{Acknowledgements}

The authors thank Hansj\"org Albrecher, Anthony Davison, Paul Embrechts, Damir
Filipovic, Christiane Lemieux and an anonymous referee for valuable feedback. As SCOR Fellow, Mathieu
Cambou thanks SCOR for financial support.

\setlength{\bibsep}{0.0pt}
\bibliographystyle{apalike}
\bibliography{CISbib}

\newpage

\begin{appendix}

\section{Direct sampling of $C^{[\lambda]}$ for shock
copulas}\label{app:AnalyticCtheta}

This appendix shows that for a certain class of shock copulas, it is possible to
directly sample from the conditional distribution $C^{[\lambda]}$, as defined
in Section~\ref{se:RejectionAlgorithm}. The Marshall--Olkin copula is a special
case of this class.

We now introduce a multivariate construction for shock copulas. Let $Z_j:\Omega\rightarrow\RR$, $j=1,\dots,m,$ for some $m\in\NN$, denote continuous independent random variables. We call the $Z_j$ ``shocks'' and denote their cdf's by $F_{Z_j}$. Suppose each component $X_j$ of $\bfX=(X_1,\dots,X_d)$ is exposed to a subset of shocks with indices $I_j\subset\{1,\dots,m\}$ through the maximum:
\begin{align}\label{eq:maxshocksvector}
(X_1,\dots,X_d)= \left(\max_{k\in I_1}Z_k,\dots,\max_{k\in I_d}Z_k\right).
\end{align}
As the $Z_j$'s are independent, the marginal cdf's $F_{X_j}$ can be calculated as
\begin{align}\label{eq:shockmarginalX}
F_{X_j}(x) = \prod_{k\in I_j}F_{Z_k}(x), \quad x\in\RR.
\end{align}
By rearranging the arguments, and due to the fact that the $Z_j$'s are independent, we can write the joint distribution of $\bfX$ as
\begin{align*}
\PP[X_1\leq x_1,\dots,X_d\leq x_d]=
\prod_{j=1}^m \PP\left[Z_j \leq \min_{k\,:\,j\in I_k} x_k \right].
\end{align*}
Hence, the copula induced by $\bfX$ is given by
\begin{align}\label{eq:shockCopula}
C(\bfu)=
\prod_{j=1}^m F_{Z_j} \left(\min_{k\,:\,j\in I_k} F_{X_k}^{-1}(u_k) \right).
\end{align}

As the copula can be expressed in terms of the independent shocks, we can write the conditional distribution $C^{[\lambda]}$ in a tractable form. To this end, let the constants $\phi_j$ and the random variables $B_j$ for $j=1,\dots,m$ be defined by
\begin{align*}
\phi_j &= \min_{k\,:\,j\in I_k} F_{X_k}^{-1}(\lambda), \quad\quad B_j= \bfone\{Z_j > \phi_j\}.
\end{align*}
Then, we can express conditioning on $\bfU \notin [0,\lambda]^d$ through the following equivalent statements
\begin{align*}
\bfU \notin [0,\lambda]^d
&\Leftrightarrow \max\{U_1,\dots,U_d\} > \lambda \\
&\Leftrightarrow X_i > F_{X_i}^{-1}(\lambda) \quad \text{ for at least one } i\in\{1,\dots,d\}\\
&\Leftrightarrow \max_{k\in I_i}Z_k > F_{X_i}^{-1}(\lambda) \quad \text{ for at least one } i\in\{1,\dots,d\}\\
&\Leftrightarrow Z_j > F_{X_k}^{-1}(\lambda) \quad \text{ for at least one } j\in\{1,\dots,m\} \text{ and }k\text{ s.t }j\in I_k\\
&\Leftrightarrow Z_j > \phi_j \quad \text{ for at least one } j\in\{1,\dots,m\}\\
&\Leftrightarrow \max\{B_1,\dots,B_m\} =1.
\end{align*}
Note that, unconditionally, the $B_j$'s are independent Bernoulli variables with parameter
\begin{align*}
p_j= \PP[B_j = 1] = 1-F_{Z_j}(\phi_j), \quad j=1,\dots,m.
\end{align*}

The following algorithm can be used to draw a realization from $C^{[\Lambda]}$. First, a realization from the 
conditional distribution of $(B_1,\dots,B_m)$ given that $\max\{B_1,\dots,B_m\} =1$ is drawn through iterative conditioning. 
Then the shocks are simulated conditionally on the $B_j$'s, which is easy as the shocks are independent under this conditioning. 
Finally, by calculating the corresponding realization of $\bfX$ with \eqref{eq:maxshocksvector}, we obtain the sample 
from $C^{[\lambda]}$. This approach is fast because the conditional distribution of $(B_1,\dots,B_m)$ given $\max \{B_1,\dots,B_m\} =1$ 
is analytically tractable, as the following algorithm also shows.
\begin{algorithm}\label{alg:directAlgorithm}
In order to draw a realization from $C^{[\lambda]}$, do:
\begin{enumerate}
\item Draw a realization from $(B_1,\dots,B_m)$ given that $\max\{B_1,\dots,B_m\} =1$ through iterative conditioning as follows.\\
For $k=1,\dots,m$:
\begin{enumerate}
\item Set
\begin{align*}
\widetilde{p}_k=
\begin{cases}
\frac{p_k}{1-\prod_{l=k}^m(1-p_l)}, & \text{ if } k=1 \text{ or } \max_{1\leq j < k} B_j = 0,\\
p_k, & \text{ if }\max_{1\leq j < k} B_j = 1.
\end{cases}
\end{align*}
\item Draw $B_k\sim \operatorname{Bernoulli}(\widetilde{p}_k$).
\end{enumerate}
\item Draw a realization from $(Z_1,\dots,Z_m)$ given $(B_1,\dots,B_m)$ as follows:\\
For $k=1,\dots,m$:
\begin{enumerate}
\item Draw $\widetilde{U}_k\sim\operatorname{U(0,1)}$
\item Set
\begin{align*}
Z_k =
\begin{cases}
F_{Z_k}^{-1}\left(p_k\widetilde{U}_k\right),         & \text{ if } B_k = 0, \\
F_{Z_k}^{-1}\left(p_k+(1-p_k)\widetilde{U}_k\right), & \text{ if } B_k = 1.
\end{cases}
\end{align*}
\end{enumerate}
\item Set $X_j=\max_{k\in I_j}Z_k$ and $U_j = F_{X_j}(X_j)$, where $F_{X_j}$ is defined in~\eqref{eq:shockmarginalX}.
\item Return $\bfU=(U_1,\dots,U_d)$.
\end{enumerate}
\end{algorithm}

Although it is not an issue for the purpose of sampling, note that for most choices of shock distributions $F_{Z_j}$, the copula C in \eqref{eq:shockCopula} does not have an analytic form. One possible choice for $F_{Z_j}$ yielding an analytic expression for $C$ is illustrated in the following example.

\begin{example}[Marshall--Olkin copula]\label{ex:MOC}
Suppose the $Z_j$'s are Fr\'echet distributed with $F_{Z_j}(x)=\exp(-s_j/x)$,
$x>0$, $j=1,\dots,m$, with scale parameters $s_j>0$. Then the $X_j$ are
also Fr\'echet distributed, with $F_{X_j}(x)=\exp(-\widetilde{s}_j/x)$,
where $\widetilde{s}_j=\sum_{k\in I_j} s_k$, $j=1,\dots,d$. The copula~\eqref{eq:shockCopula} then reduces to
\begin{align*}
C(\bfu)&=\prod_{j=1}^m \exp\left\{-s_j\left(\min_{i:j\in I_i}\left(\frac{-\log u_i}{\widetilde{s}_i}\right)^{-1}\right)^{-1}\right\}= \prod_{j=1}^m \min_{i:j\in I_i} u_i^{s_j/\widetilde{s}_i}.
\end{align*}
This copula is of the Marshall--Olkin type, see \cite{MarshallOlkin1967}. As an example, consider $d=2$, $m=3$, $I_1=\{1,3\}$, and $I_2=\{2,3\}$. In this case, $X_1 = \max\{Z_1,Z_3\}$, $X_2 = \max\{Z_2,Z_3\}$, and the copula can be written as
\begin{align*}
C(u_1,u_2)
= u_1^{s_1/(s_1+s_3)} \cdot u_2^{s_2/(s_2+s_3)} \cdot
\min\left\{u_1^{s_3/(s_1+s_3)},u_2^{s_3/(s_2+s_3)}\right\}
= \min\left\{u_1 u_2^{s_2/(s_2+s_3)},u_2 u_1^{s_1/(s_1+s_3)}\right\}.
\end{align*}
\end{example}

\end{appendix}

\end{document}